\documentclass[twocolumn,preprint]{aastex631}

\usepackage{amsmath,amstext}
\usepackage[T1]{fontenc}
\usepackage{apjfonts} 
\usepackage{natbib}
\usepackage{longtable}
\usepackage{gensymb}
\usepackage{microtype,longtable,verbatim,float,booktabs,graphicx}

\newcommand{\target}{RM160}
\newcommand{\targetLong}{SDSS J141041.25+531849.0}
\newcommand{\angstrom}{\mbox{\normalfont\AA}}
\newcommand{\MgII}{MgII}

\newcommand{\Ha}{\hbox{{\rm H}$\alpha$}}
\newcommand{\Hg}{\hbox{{\rm H}$\gamma$}}
\newcommand{\HavNII}{\hbox{({\rm H}$\alpha$+{\rm [N}\kern 0.1em{\sc II}{\rm ]})}}
\newcommand{\Hb}{\hbox{{\rm H}$\beta$}}
\newcommand{\SII}{\hbox{{\rm [S}\kern 0.1em{\sc ii}{\rm ]}}}
\newcommand{\NII}{\hbox{{\rm [N}\kern 0.1em{\sc ii}{\rm ]}}}
\newcommand{\OII}{\hbox{{\rm [O}\kern 0.1em{\sc II}{\rm ]}}}
\newcommand{\OIII}{\hbox{{\rm [O}\kern 0.1em{\sc iii}{\rm ]}}}
\newcommand{\NeIII}{\hbox{{\rm [Ne}\kern 0.1em{\sc III}{\rm ]}}} \newcommand{\NeIIIvHeI}{\hbox{{\rm [Ne}\kern 0.1em{\sc III}{\rm ]}+{\rm [He}\kern 0.1em{\sc I}{\rm ]}}}
\newcommand{\NeV}{\hbox{{\rm [Ne}\kern 0.1em{\sc v}{\rm ]}}}
\newcommand{\HeII}{\hbox{{\rm He}\kern 0.1em{\sc II}}}
\newcommand{\HeI}{\hbox{{\rm He}\kern 0.1em{\sc I}}}
\newcommand{\HII}{\hbox{{\rm H}\kern 0.1em{\sc II}}}

\newcommand{\emissAll}{\MgII, \Hb, and \Ha}

\def\arrvline{\hfil\kern\arraycolsep\vline\kern-\arraycolsep\hfilneg}

\newcommand{\GrierHbLagObs}{$\tau_{\rm{H\beta,} \rm{obs}} = 31.3^{+8.1}_{-4.1}$ days}
\newcommand{\GrierHaLagObs}{$\tau_{\rm{H\alpha,} \rm{obs}} = 27.7^{+5.3}_{-4.7}$ days}
\newcommand{\HomayouniMgIILagObs}{$\tau_{\rm{MgII,} \rm{obs}} = 144.7^{+24.7}_{-22.6}$ days}

\newcommand{\HbLagObsNum}{$31.3^{+8.1}_{-4.1}$ days}
\newcommand{\HaLagObsNum}{$27.7^{+5.3}_{-4.7}$ days}
\newcommand{\MgIILagObsNum}{$144.7^{+24.7}_{-22.6}$ days}

\newcommand{\targetBHMass}{$(M_{\rm BH}/10^{7}M_\odot) = 7.0^{+1.7}_{-1.3}$}

\newcommand{\cosmo}{$\Lambda$CDM cosmology with $\Omega_{\Lambda}$ = 0.7, $\Omega_{M}$ = 0.3, and $H_{0}$ = 70~km~s$^{-1}$~Mpc$^{-1}$}

\newcommand{\PyROA}{\texttt{PyROA}}
\newcommand{\JAV}{\texttt{JAVELIN}}
\newcommand{\MEM}{\texttt{MEMECHO}}

\newcommand{\CARAMEL}{\texttt{CARAMEL}}
\newcommand{\BRAINS}{\texttt{BRAINS}}

\shorttitle{Resolving the BLR of RM160}
\shortauthors{Fries et al.}

\begin{document}

\defcitealias{Fries2023}{F23}

% \title{\large \bf The SDSS-V Black Hole Mapper Reverberation Mapping Project: Resolving the Broad-Line Region of \target\ with Velocity-Resolved Reverberation Mapping}

\title{\large \bf The SDSS-V Black Hole Mapper Reverberation Mapping Project: A Kinematically Variable Broad-Line Region and Consequences for Masses of Luminous Quasars}

\correspondingauthor{Logan Fries}
\email{logan.fries@uconn.edu}

\author[0000-0001-8032-2971]{Logan B. Fries}
\affil{Department of Physics, 196A Auditorium Road, Unit 3046, University of Connecticut, Storrs, CT 06269, USA}

\author[0000-0002-1410-0470]{Jonathan R. Trump}
\affil{Department of Physics, 196A Auditorium Road, Unit 3046, University of Connecticut, Storrs, CT 06269, USA}

% \author{The Team}

\author[0000-0003-1728-0304]{Keith Horne}
\affiliation{SUPA Physics and Astronomy, University of St. Andrews, Fife, KY16 9SS, Scotland, UK}

\author[0000-0001-9776-9227]{Megan C. Davis}
\affil{Department of Physics, 196A Auditorium Road, Unit 3046, University of Connecticut, Storrs, CT 06269, USA}

\author[0000-0001-9920-6057]{Catherine~J.~Grier}
\affiliation{Department of Astronomy, University of Wisconsin-Madison, Madison, WI 53706, USA} 

\author[0000-0003-1659-7035]{Yue Shen}
\affiliation{Department of Astronomy, University of Illinois at Urbana-Champaign, Urbana, IL 61801, USA}
\affiliation{National Center for Supercomputing Applications, University of Illinois at Urbana-Champaign, Urbana, IL 61801, USA}

\author[0000-0002-6404-9562]{Scott F. Anderson}
\affiliation{Astronomy Department, University of Washington, Box 351580, Seattle, WA 98195, USA}

\author[0000-0002-4459-9233]{Tom Dwelly}
\affiliation{Max-Planck-Institut f{\"u}r extraterrestrische Physik, Giessenbachstra\ss{}e, 85748 Garching, Germany}

\author[0000-0002-0957-7151]{Y. Homayouni}
\affiliation{Department of Astronomy \& Astrophysics, 525 Davey Lab, The Pennsylvania State University, University Park, PA 16802, USA}
\affiliation{Institute for Gravitation and the Cosmos, The Pennsylvania State University, University Park, PA 16802, USA}

\author[0000-0002-6770-2627]{Sean Morrison}
\affiliation{Department of Astronomy, University of Illinois at Urbana-Champaign, Urbana, IL 61801, USA}

\author[0000-0001-8557-2822]{Jessie C. Runnoe}
\affiliation{Department of Physics and Astronomy, Vanderbilt University, Nashville, TN 37235, USA}

\author[0000-0002-3683-7297]{Benny Trakhtenbrot}
\affiliation{School of Physics and Astronomy, Tel Aviv University, Tel Aviv 69978, Israel}

% % Project 77 (alphabetically)

\author[0000-0002-9508-3667]{Roberto J. Assef}
\affiliation{Instituto de Estudios Astrof\'isicos, Facultad de Ingeniera\'ia y Ciencias, Universidad Diego Portales, Av. Ej\'ercito Libertador 441, Santiago, Chile 8370191}

\author[0000-0002-3601-133X]{Dmitry Bizyaev}
\affiliation{Apache Point Observatory and New Mexico State
University, P.O. Box 59, Sunspot, NM, 88349-0059, USA}
\affiliation{Sternberg Astronomical Institute, Moscow State
University, Moscow}

\author[0000-0002-0167-2453]{W. N. Brandt}
\affiliation{Department of Astronomy \& Astrophysics, 525 Davey Lab, The Pennsylvania State University, University Park, PA 16802, USA}
\affiliation{Institute for Gravitation and the Cosmos, The Pennsylvania State University, University Park, PA 16802, USA}
\affiliation{Department of Physics, 104 Davey Lab, The Pennsylvania State University, University Park, PA 16802, USA}

\author[0000-0003-1317-8847]{Peter Breiding}
\affiliation{The William H. Miller III Department of Physics \& Astronomy, Johns Hopkins University, Baltimore, MD 21218, USA}

\author[0000-0002-8725-1069]{Joel Brownstein}
\affiliation{Department of Physics and Astronomy, University of Utah, 115 S. 1400 E., Salt Lake City, UT 84112, USA}

\author[0000-0002-4469-2518]{Priyanka Chakraborty}
\affiliation{Center for Astrophysics | Harvard \& Smithsonian, Cambridge, MA 02138, USA}

\author[0000-0002-1763-5825]{P. B. Hall}
\affiliation{Department of Physics and Astronomy, York University, Toronto, ON M3J 1P3, Canada}

\author[0000-0002-6610-2048]{Anton M. Koekemoer}
\affiliation{Space Telescope Science Institute, 3700 San Martin Dr., Baltimore, MD 21218, USA}

\author[0000-0002-9790-6313]{H\'ector J. Ibarra-Medel}
\affiliation{Universidad Nacional Aut\'onoma de M\'exico, Instituto de Astronom\'ia, AP 70-264, CDMX 04510, Mexico}

\author[0000-0002-7843-7689]{Mary Loli Martínez-Aldama}
\affiliation{Astronomy Department, Universidad de Concepción, Casilla 160-C, Concepción 4030000, Chile}

\author[0000-0002-1656-827X]{C. Alenka Negrete}
\affiliation{CONAHCyT Research Fellow - Universidad Nacional Aut\'onoma de M\'exico, Instituto de Astronom\'ia, AP 70-264, CDMX 04510, Mexico}

\author[0000-0002-2835-2556]{Kaike Pan}
\affiliation{Apache Point Observatory and New Mexico State
University, P.O. Box 59, Sunspot, NM, 88349-0059, USA}

\author[0000-0001-5231-2645]{Claudio Ricci}
\affiliation{Instituto de Estudios Astrof\'isicos, Facultad de Ingenier\'ia y Ciencias, Universidad Diego Portales, Av. Ej\'ercito Libertador 441, Santiago, Chile}
\affiliation{Kavli Institute for Astronomy and Astrophysics, Peking University, Beijing 100871, People's Republic of China}

\author[0000-0001-7240-7449]{Donald P.\ Schneider} 
\affiliation{Department of Astronomy \& Astrophysics, 525 Davey Lab, The Pennsylvania State University, University Park, PA 16802, USA}
\affiliation{Institute for Gravitation and the Cosmos, The Pennsylvania State University, University Park, PA 16802, USA}

\author[0000-0001-9616-1789]{Hugh W. Sharp}
\affiliation{Department of Physics, 196A Auditorium Road, Unit 3046, University of Connecticut, Storrs, CT 06269, USA}

\author[0009-0003-8591-0061]{Theodore B. Smith}
\affiliation{Department of Physics, 196A Auditorium Road, Unit 3046, University of Connecticut, Storrs, CT 06269, USA}

\author[0000-0002-8501-3518]{Zachary Stone}
\affiliation{Department of Astronomy, University of Illinois at Urbana-Champaign, Urbana, IL 61801, USA}

\author[0000-0001-8433-550X]{Matthew J. Temple}
\affil{Instituto de Estudios Astrof\'isicos, Facultad de Ingenier\'ia y Ciencias, Universidad Diego Portales, Av. Ej\'ercito Libertador 441, Santiago, Chile}

\begin{abstract}
We present a velocity-resolved reverberation mapping analysis of the hypervariable quasar \target\ (\targetLong) at $z = 0.359$ with 153 spectroscopic epochs of data representing a ten-year baseline (2013-2023). We split the baseline into two regimes based on the 3x flux increase in the light curve: a `low state' phase during the years 2013-2019 and a `high state' phase during the years 2022-2023. The velocity-resolved lag profiles (VRLP) indicate that gas with different kinematics dominates the line emission in different states. The \Hb\ VRLP begins with a signature of inflow onto the BLR in the `low state', while in the `high state' it is flatter with less signature of inflow. The \Ha\ VRLP begins consistent with a virialized BLR in the `low state', while in the `high state' shows a signature of inflow. The differences in the kinematics between the Balmer lines and between the `low state' and the `high state' suggests complex BLR dynamics. We find that the BLR radius and velocity (both FWHM and $\sigma$) do not obey a constant virial product throughout the monitoring period. We find that BLR lags and continuum luminosity are correlated, consistent with rapid response of the BLR gas to the illuminating continuum. The BLR kinematic profile changes in unpredictable ways that are not related to continuum changes and reverberation lag. Our observations indicate that non-virial kinematics can significantly contribute to observed line profiles, suggesting caution for black-hole mass estimation in luminous and highly varying quasars like \target.
\end{abstract}

\keywords{Quasars -- Broad Line Kinematics -- Galaxy Evolution -- Active Galactic Nuclei}
  
\section{Introduction}
The optical spectra of Active Galactic Nuclei (AGN) are characterized by a blue continuum and %the existence of 
broad emission lines (e.g., \citealp{Seyfert1943}). Broad emission lines are created when the continuum flux photoionizes a distribution of gas called the broad line region (BLR). Since the BLR is located close to the central engine (black hole + accretion disk) and is assumed to be moving under orbits dominated by gravity (i.e., virialized; \citealp{Peterson2004, Bentz2009, Grier2013}), the emission lines are Doppler broadened by the high velocity gas. 

In \cite{Fries2023}, we presented an analysis of dramatic variability of three broad emission lines (\MgII, \Hb, and \Ha) in the quasar \targetLong\ (hereafter \target). This quasar was observed to have normal ``line breathing'' behavior (i.e., the broad emission-line widths decrease with an increase in continuum flux and vice versa), but also exhibited dramatic broad emission-line radial velocity variations, ranging from $\Delta{v}$ $\sim$800~km~s$^{-1}$ to $\sim$1600~km~s$^{-1}$, relative to the systemic redshift. The radial velocity variations are consistent with a multi-faceted explanation: (1) a bulk inflow of the BLR gas with a gradient of higher velocity at smaller radii, (2) an azimuthal asymmetry circulating around in the inner regions of the BLR, and (3) stochastic flux-driven changes to the optimal emission region (i.e., ``line breathing'') \citep{Barth2015, Wang2020}. We presented this phenomenological model with the aim of further investigating the structure and kinematics of the BLR with reverberation mapping \citep{Blandford1982, Peterson1993, Cackett2021}.

Since the BLR is on the order of tens to hundreds of light-days in size, there are only a handful of studies that have resolved the BLR spatially using interferometric techniques (e.g., \citealp{GravityCollab2020}), but these are only possible for low-redshift AGN. Since we cannot spatially resolve the BLR in \textit{most} quasars, we must turn to other techniques. Reverberation mapping is a technique that trades spatial resolution for temporal resolution. The basic idea of reverberation mapping is that the flux variations in the broad emission lines closely follow (i.e., lag) the stochastic flux variations from the central engine. The lag ($\tau$) between these two signals corresponds to the light-travel time for ionizing radiation emitted by the central engine to reach the BLR. The lag between the central engine and an arbitrary emission line corresponds to the ``response-weighted radius'' (or ``optimal emission radius'') of the given emission line using $R_{\rm BLR} = c \tau$. Probing different emission lines via reverberation mapping allows one to infer the structure of the BLR by understanding where different emission lines are most ``optimally'' emitted. Indeed, it has been found that higher-ionization lines respond to continuum variations more rapidly, indicating that the BLR is radially stratified \citep{Gaskell1986, Peterson1993, Bischoff1999, Kollatschny2018}. 

Assuming a linear response of a given emission line to continuum photons, the responding emission-line light curve can be described by the convolution of the continuum light curve with a transfer function \citep{Peterson1993}:

\begin{equation}
    \label{ReverberationTransferFunction}
    \Delta L \, (v, \, t) = \int_{-\infty}^{\infty} \; \Psi(v, \, \tau) \; \Delta C(t - \tau) \; d\tau
\end{equation}
where $\Delta L \, (v, \, t)$ is the change in emission-line luminosity as a function of line-of-sight velocity $v$ at time $t$ relative to reference levels (e.g., the light curve mean or median), $\Delta C(t - \tau)$ is the change in continuum light curve relative to reference levels, and $\Psi(v, \, \tau)$ is the transfer function, or velocity-delay map, which describes the response of the emission line (after some time $\tau$) to the continuum as a function of line-of-sight velocity, $v$. Unraveling the transfer function (i.e., recovering the velocity-delay map), $\Psi(v, \, \tau)$, is the primary goal of reverberation mapping as the kinematics, geometry, and physics of the BLR are encoded within it \citep{Horne2004}. 

The aforementioned reverberation mapping techniques involve measuring the lag across the entire emission line (i.e., the integrated lag), but it is also possible to measure the lag in different velocity bins. This idea of measuring the lag as a function of line-of-sight velocity, $v$, is called velocity-resolved reverberation mapping and allows one to make inferences about the kinematics of the BLR. Velocity-resolved reverberation mapping has unveiled a diversity of kinematics in the BLR (e.g., \citealp{Denney2009a}). For a simple BLR geometry, a virialized BLR will have longer lags in the bins at the line center, while having the shortest lags in the high-velocity wings. This is consistent with a virialized BLR interpretation since the lowest velocity gas should be further out. An infalling BLR will have the longest lags in the blueshifted bins and the shortest lags in the redshifted bins. This means that the gas on the far side (highest lags) is moving toward us (blueshifted) and the gas on the near side (shortest lags) is moving away from us (redshifted). An outflowing BLR will have the longest lags in the redshifted bins and the shortest lags in the blueshifted bins. This means that the gas on the far side (longest lags) is moving away from us (redshifted) and the gas on the near side (shortest lags) is moving toward us (blueshifted). Another signature that arises from velocity-resolved reverberation mapping is the presence of an `M'-shape in the lags as a function of line-of-sight velocity. This `M'-shape has been found in various emission lines previously such as \Hb, \Hg, and Ly$\alpha$ \citep{Pei2017, Horne2021, KaiXing2022}. This `M'-shaped structure can be interpreted as a flat disk or spherical BLR where the illuminating source is emitting isotropically \citep{Pei2017, KaiXing2022} or an inclined Keplerian disk \citep{Horne2021}. Recent papers use dynamical models to infer additional details of the BLR kinematics and geometry from velocity-resolved reverberation mapping (\citealp{Villafana2022}, Stone et al. in prep).

Velocity-resolved reverberation mapping requires a dataset that has high signal-to-noise, high cadence, high spectral resolution, and a lengthy observation campaign. As such, the AGN that have had this analysis applied to them has been limited to a small sample of nearby Seyfert-1 AGN \citep{Denney2009b, Bentz2009, Bentz2010a, Barth2011, Grier2013, DeRosa2015, Du2016, DeRosa2018, Bentz2021, U2022, Li2022, KaiXing2022}. These previous efforts were single-target based, whereas the multi-object, long-duration, and many-epoch spectroscopic monitoring campaign of the Sloan Digital Sky Survey (SDSS) Black Hole Mapper Reverberation Mapping (BHM-RM) Project represents an opportunity to dramatically expand this analysis to a large set of quasars.

Using simple models (i.e., measuring the lag as a function of line of sight velocity) can illuminate inferences about the kinematics of the BLR (e.g., \citealt{U2022}), but in order to get a more complete view of the structure and kinematics of the BLR one must model it. There are two independent approaches for this kind of analysis: (1) through forward modeling (\CARAMEL, \citealp{Pancoast2011, Pancoast2014} or \BRAINS, \citealp{Li2013, Li2018}), which uses a set of self-consistent models to explore the parameter space to find the solution that best matches the data or (2) through an inverse problem (\MEM, \citealp{Horne1994}), which extracts the transfer function directly from the data. \MEM\ has been used to model a number of previous objects (\citealp{Bentz2010b, Grier2013, Xiao2018, Horne2021}), and it finds diverse kinematic signatures of the BLR gas. In future work, Stone et al. in prep will show full dynamical modeling of \target\ using \BRAINS.

In this work we apply velocity-resolved reverberation mapping techniques to the luminous quasar \target\ to evaluate the phenomenological model proposed in \cite{Fries2023} and gain further insight into its kinematics and structure. Section~\ref{Sec2} describes the properties of RM160, the spectroscopic observations, and the photometric observations used in this study. Section~\ref{Sec3} describes the process of measuring the integrated time delays. Section~\ref{Sec4} describes the process of measuring the velocity-resolved time delays. Section~\ref{Sec5} describes our interpretation of the data. Finally, Section~\ref{Sec6} summarizes our results.

\begin{figure*}[t]%[ht!]
\epsscale{1.1}
\plotone{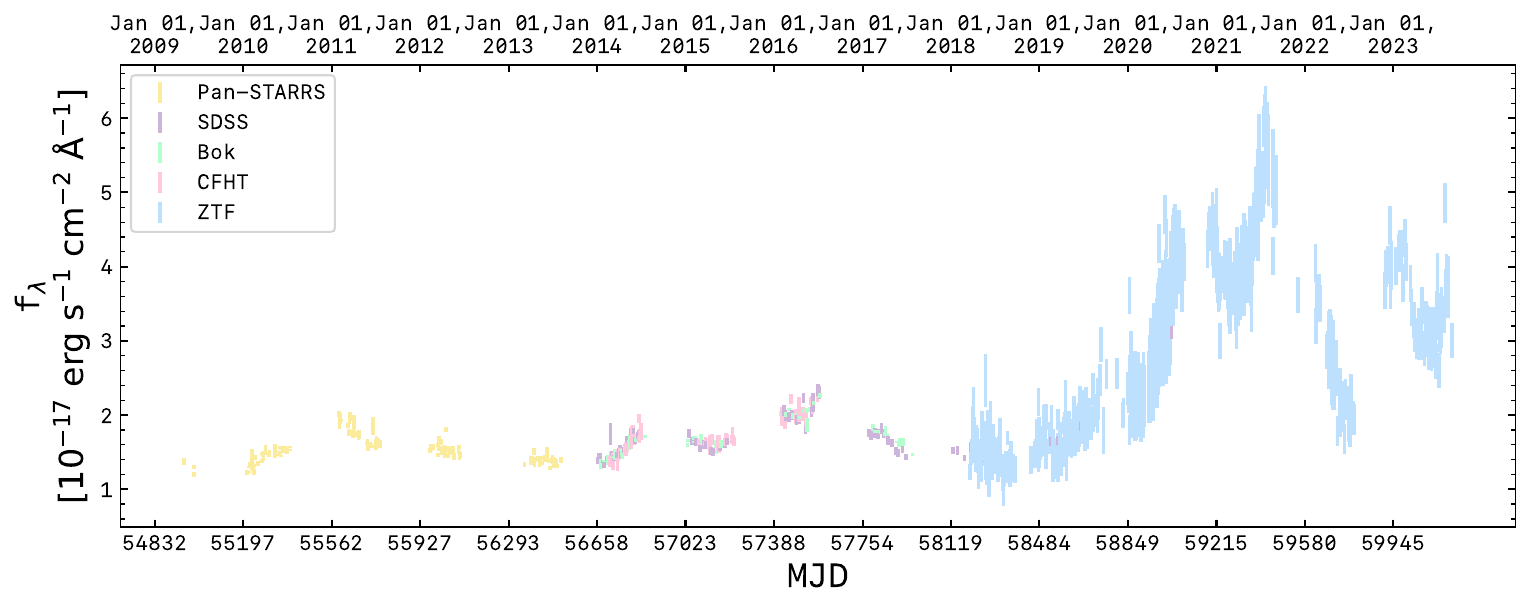}
\figcaption{Photometric light curve for \target, demonstrating the long baseline of our photometric coverage. The photometric (continuum) light curve is comprised of data from Bok (green points), CFHT (pink points), Pan-STARRS (yellow points), SDSS (purple points), and ZTF (blue points). These data include $g$, $r$, $i$ photometry and been merged together to account for instrumental differences using \texttt{PyCali} \citep{Li2014}.
\label{Fig:photometric_light_curve}}
\end{figure*}

Throughout this work, we assume a \cosmo.

\section{Observations}
\label{Sec2}

\subsection{Properties of \target}
\label{Sec:TargetProperties}
\target\ is a luminous quasar located at R.A. = $14^{\rm h} 10^{\rm m} 41^{\rm s}.2$ and Dec = $53^{\circ}18'48.995''$(J2000) with $z$ = 0.359. There are previously measured reverberation mapping lags for \target. \cite{Grier2017b} found an observed-frame \Hb\ lag of \GrierHbLagObs\ and an observed-frame \Ha\ lag of \GrierHaLagObs\ using the 2014 spectroscopic data of \target. \cite{Homayouni2020} found an observed-frame \MgII\ lag of \HomayouniMgIILagObs\ for \target\ using the 2014-2017 spectroscopic data. \cite{Grier2017b} also computed a black hole mass of \targetBHMass\ using the \Hb\ reverberation mapping results. We expand to 10 years of spectroscopic monitoring and go beyond these integrated lags to perform velocity-resolved reverberation mapping.

\subsection{BHM-RM/SDSS-RM Spectroscopy}
\label{Sec:Spectroscopy}
The spectroscopic data were taken by the Sloan Digital Sky Survey Reverberation Mapping (SDSS-RM) project from 2014-2020 \citep{Shen2015, Shen2019} and Black Hole Mapper Reverberation Mapping (BHM-RM) project from 2021-present (Trump et al. in prep). The SDSS-RM survey was a part of the third \citep{Eisenstein2011} and fourth \citep{Blanton2017} iterations of the Sloan Digital Sky Survey (SDSS, \citealp{York2000}). The BHM-RM survey is a part of the fifth iteration \citep{Kollmeier2019} of SDSS.

The data from both SDSS-RM and BHM-RM were taken using the plate-based, fiber-fed SDSS BOSS spectrograph \citep{Smee2013} and the robotic focal plane system (FPS) based spectrograph \citep{Sayres2022} on the 2.5m SDSS telescope \citep{Gunn2006} at Apache Point Observatory in Sunspot, New Mexico. The plate-based data were taken from 2014-2021, while the FPS data were taken from 2022-present. There are irregularities between the cadences for a given year due to weather, SDSS scheduling priorities, instrument upgrades between successive generations of SDSS, and issues arising from the COVID-19 pandemic. There is a difference in spectrophotometric precision between SDSS-RM (3\%, \citealp{Shen2015, Shen2019}) and BHM-RM (5\%, Trump et al. in prep). The SDSS-RM (SDSS-III and SDSS-IV) data were reduced using the v5\_13\_0 version of idlspec2d \citep{Bolton2012}, which is the BOSS spectroscopic reduction pipeline. The BHM-RM data were reduced using the v6\_0\_9 version of idlspec2d (Morrison et al. in prep). 

In order to obtain light curves of the \MgII, \Hb, and \Ha\ emission lines, we integrated the flux contained within each line after subtracting an (epoch-dependent) local continuum, as described in detail in Section~3.4 of \cite{Fries2023}. We also obtained line widths of the RMS spectra of each emission line using the first moment of the line profile, which is also described in Section~3.4 of \cite{Fries2023}.

\subsection{Photometry}
\label{Sec:Photometry}

Our continuum light curve consists of \textit{g} and \textit{i} band photometry from the Steward Observatory's 2.3m Bok telescope and the 3.6m Canada-France-Hawaii Telescope (CFHT), which span from 2014-2017 (see \citealp{Kinemuchi2020} for more details regarding the photometry from Bok and CFHT). Additionally, it consists of \textit{g}, \textit{r}, and \textit{i} band photometry from Pan-STARRS from 2010-2013 \citep{Flewlling2020} and from the Zwicky Transient Facility (ZTF, \citealp{Masci2019}) from 2018-2023. It also includes synthetic SDSS \textit{r} band photometry from 2014-2019. The aforementioned photometric observations were merged together using the public code \texttt{PyCali} \citep{Li2014} to account for instrumental differences. For more information on the merging process see Section~3.2 of \citet{Shen2024}. Figure~\ref{Fig:photometric_light_curve} shows the continuum light curve for \target.

\section{Integrated Lags}
\label{Sec3}
\subsection{Lag Measurement Methodology}
\label{Sec:LagMeasuringMethod}
We use the photometric light curve as the driving light curve (continuum) in our reverberation mapping analysis and the emission-line light curves as the responding light curves. We use \PyROA\footnote{https://github.com/FergusDonnan/PyROA} \citep{Donnan2021} to fit the light curves in order to recover the emission-line lags. \PyROA\ uses a Bayesian MCMC approach to model quasar variability with a running optimal average (ROA), where the optimal average is a weighted average with the weights being the inverse variance. The ROA is applied within a window function whose width can be specified. The window function acts as a filter whereby the impact of data points far from the point of interest are reduced. For a continuous monitoring program (i.e., data with no seasonal gaps) a smaller window function width is optimal, whereas for a monitoring program like SDSS-RM/BHM-RM with seasonal gaps, a larger window function width is optimal.

\PyROA\ offers unique functionality compared to more traditional lag measuring methods. In particular, \PyROA\ lets the user fit multiple responding light curves at once (where the driving light curve is constrained with respect to all of the responding light curves), allows the user the ability to choose different shapes for the delay distribution, or transfer function, in the fit. Additionally, \PyROA\ offers the user the ability to input an extra error rescaling parameter ($\sigma$), which has been demonstrated to recover lags better than \JAV\ when the flux errors are underestimated \citep{Donnan2021}. \PyROA\ has been applied recently in measuring accretion disk lags in PG 1119+120 \citep{Donnan2023} and Mrk~817 \citep{Cackett2023} as well as broad line lags for a catalog of 849 SDSS-RM broad-line quasars \citep{Shen2024}. 

\begin{figure*}[t!]%[ht!]
\epsscale{1.1}
\plotone{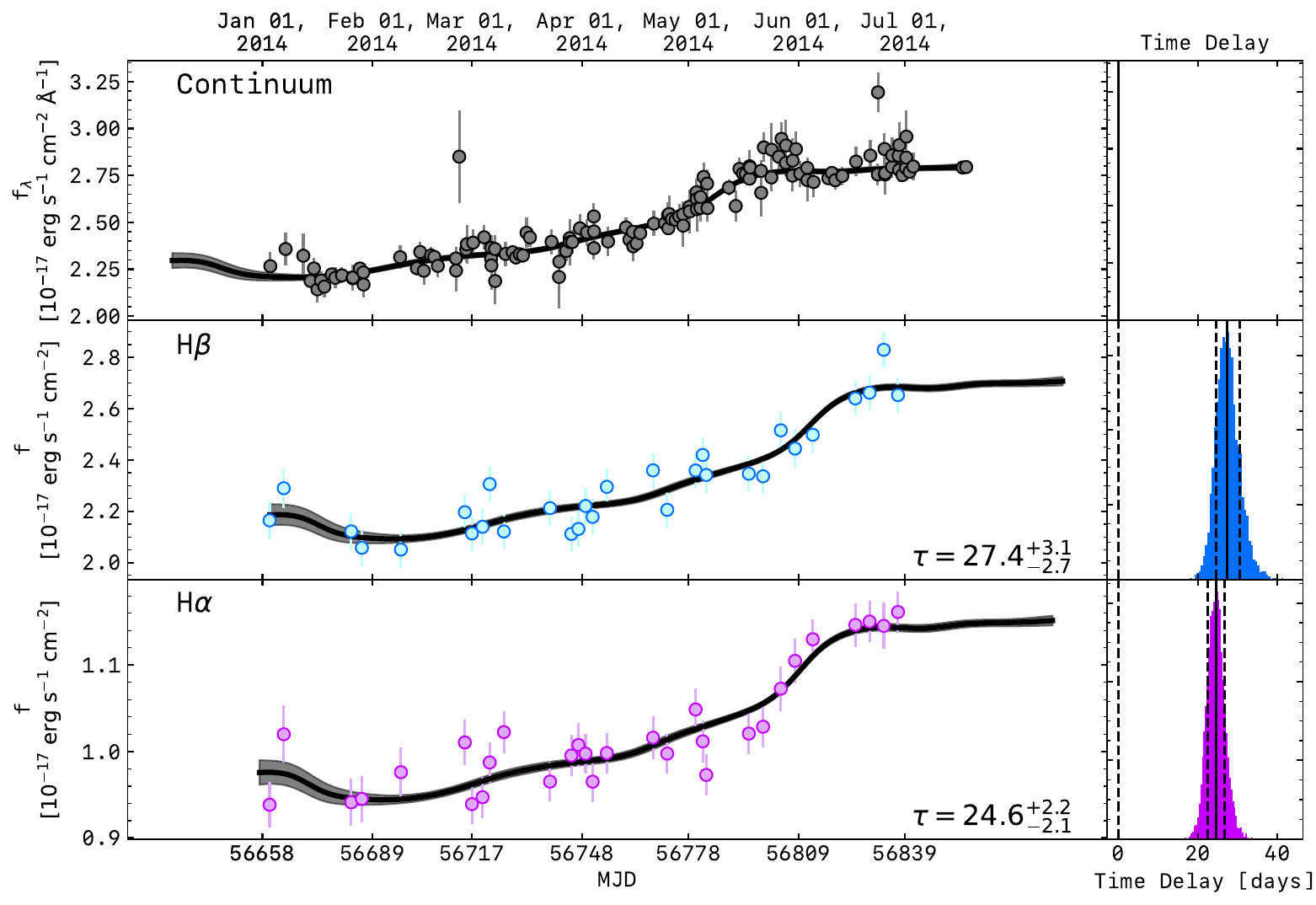}
\figcaption{2014 Balmer light curves with best-fit \PyROA\ model overlaid as a solid black line with the corresponding error envelope. The top, middle and bottom panels show the continuum, \Hb, and \Ha\ light curves respectively. The best-fit time delay along with its corresponding errors are on the bottom right of the \Hb\ and \Ha\ light curve panels. The posterior distribution for the time delay is on the right panel of each responding light curve where the median value and 68$^{\rm th}$ percent errors are the solid and dashed lines, respectively. We also show, on the right hand panels, a vertical dashed line at the zero time delay for reference. We find an \Hb\ time delay of $27.4^{+3.1}_{-2.7}$~days and an \Ha\ time delay of $24.6^{+2.2}_{-2.1}$~days, which are consistent with the \JAV\ time delays computed in \cite{Grier2017b}.
\label{Fig:PyROA_2014_Balmer}}
\end{figure*}

\PyROA\ lets the user input lower and upper limits to uniform priors, which are hard limits, on 5 parameters: $A$ (RMS flux of each light curve, in units of RMS flux to the MAD light curve), $B$ (mean flux of each light curve, in units of mean flux to the median light curve), $\tau$ (the time delay of each light curve, in units of days), $\Delta$ (the width of the window function of the ROA algorithm, in units of days), and $\sigma$ (an extra error rescaling parameter, in units of flux). Our choice in the lower and upper limits to the uniform priors is listed below\footnote{We note that the way the $A$ and $B$ priors are computed have changed from \PyROA\ v3.1.0 to v3.2.0. For clarity, we have done this analysis on \PyROA\ v3.2.0.}:

\begin{itemize}
    \item $A$: [0.0, 2.0] 
    \item $B$: [0.0, 2.0] 
    \item $\tau$: [0.0, 300.0]
    \item $\Delta$: [5.0, 50.0]
    \item $\sigma$: [0.0, 10.0]
\end{itemize}

\begin{figure*}[t!]%[ht!]
\epsscale{1.1}
\plotone{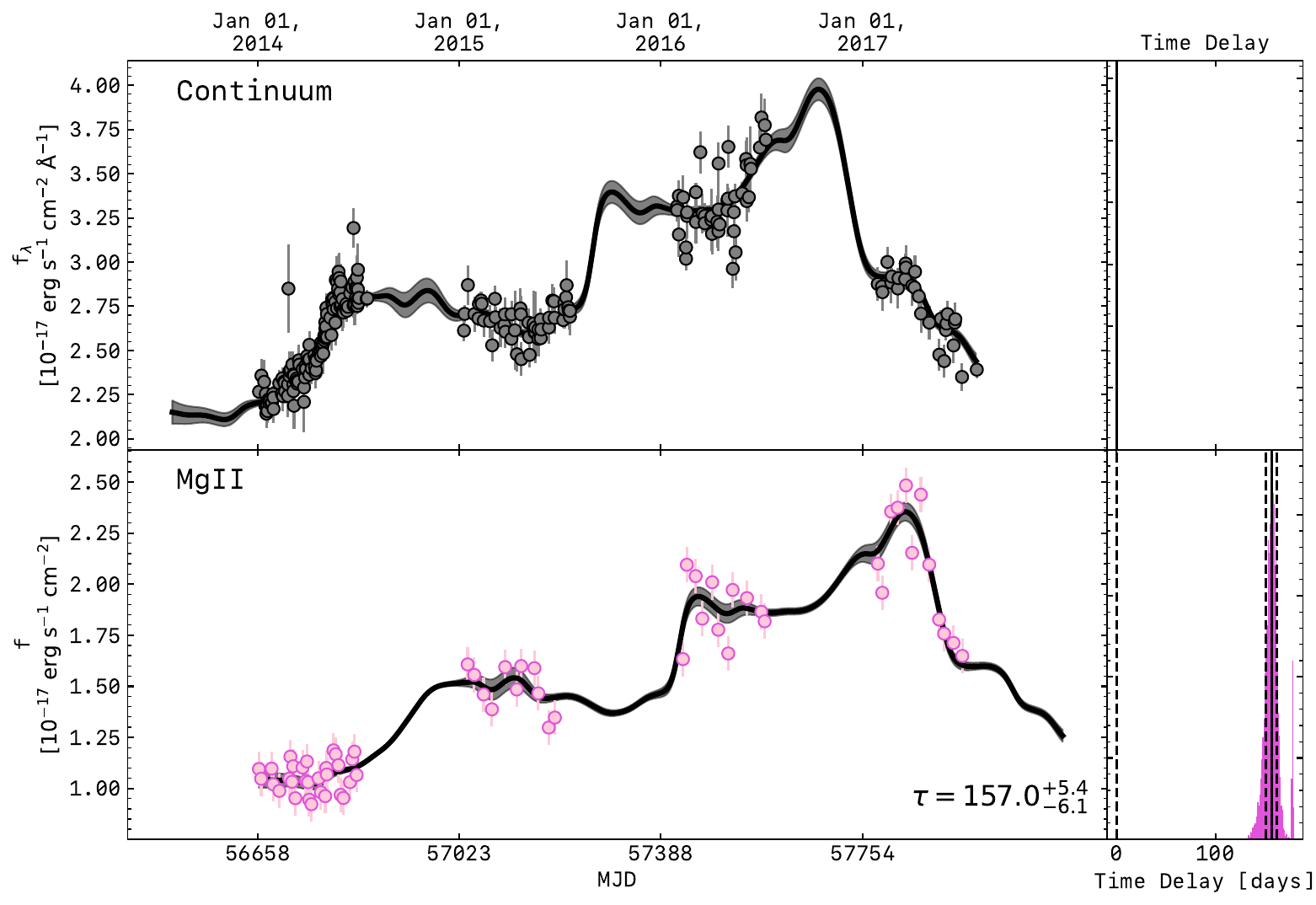}
\figcaption{2014-2017 \MgII\ light curve with best-fit \PyROA\ model overlaid as a solid black line with the corresponding error envelope. The top panel shows the continuum, while the bottom panel shows the \MgII\ light curve. The best-fit time delay along with its corresponding errors are on the bottom right of the \MgII\ light curve panel. The posterior distribution for the time delay is on the right panel of each responding light curve where the median value and 68$^{\rm th}$ percent errors are the solid and dashed lines, respectively. We also show, on the right hand panels, a vertical dashed line at the zero time delay for reference. We find an \MgII\ time delay of $157.0^{+5.4}_{-6.1}$~days, which is consistent with the \JAV\ time delay computed in \cite{Homayouni2020}.
\label{Fig:PyROA_2014_2017_MgII}}
\end{figure*}

For our \PyROA\ fits, we first scale the responding light curves' flux to be on the same order of magnitude as the driving light curve, as this causes the fit to converge more quickly. We use a delta-function delay distribution shape. We find that the inclusion of an additional free parameter (i.e., more complex shapes of the delay distribution such as a Gaussian or log-Gaussian) leads to more degeneracies in the fit, which leads to larger uncertainties in the best-fit lags. Finally, we use 25,000 MCMC samples and discard the first 20,000 of them for burn-in in our fits.

\subsection{Comparison to Previously Measured Lags}
\label{Sec:JavelinComparison}
In order to motivate our usage of \PyROA, we compare the results of \PyROA\ with \JAV. The time delays of the BLR for \target\ have been studied previously \citep{Grier2017b, Homayouni2020} on smaller time scales. These studies used \JAV, which models quasar variability with a damped random walk (DRW) model with a top-hat transfer function shape. \PyROA\ models quasar variability with a ROA algorithm, assumes no shape to the driving light curve, and allows the user to supply a range of transfer function shapes (delta function, Gaussian, log Gaussian, truncated Gaussian, etc). Futhermore, \PyROA\ does a better job at dealing with seasonal gaps and aliasing over long baselines (see Section 4.2 in \citealt{Shen2024}). For more detail on the usage of \JAV\ as it pertains to \target, please see \cite{Grier2017b} and \cite{Homayouni2020}.

Using the emission line light curves from the monitoring periods of \cite{Grier2017b} and \cite{Homayouni2020}, we use \PyROA\ to measure the lags for the Balmer (\Hb\ and \Ha) lines and MgII, respectively. Since \PyROA\ allows for multiple responding light curves to be fit at once, we fit \Hb\ and \Ha\ simultaneously. Our resultant fit from \PyROA\ for the Balmer lines is shown in Fig.~\ref{Fig:PyROA_2014_Balmer}. We find an observed-frame \Hb\ lag of $27.4^{+3.1}_{-2.7}$~days and an observed-frame \Ha\ lag of $24.6^{+2.2}_{-2.1}$~days. This is consistent with the observed-frame \Hb\ and \Ha\ lags from \cite{Grier2017b} using \JAV\ of \HbLagObsNum\ and \HaLagObsNum, respectively. Our resultant fit from \PyROA\ for \MgII\ is shown in Fig.~\ref{Fig:PyROA_2014_2017_MgII}. We find an observed-frame \MgII\ lag of $157.0^{+5.4}_{-6.1}$~days. This is consistent with the observed-frame \MgII\ lag from \cite{Homayouni2020} using \JAV\ of \MgIILagObsNum. Table~\ref{Tab:JAV_Comparison} shows the comparison between \PyROA\ and \JAV. We find that the time delays measured by \PyROA\ and \JAV\ agree within the uncertainties validating our use of \PyROA\ for measuring reverberation lags in this work. The \PyROA\ errors are slightly smaller than \JAV. We anticipate this as due to a more well constrained fit by using multiple responding light curves to constrain the fit. Further discussion of \PyROA\ and \JAV\ error comparison is given in \cite{Donnan2021}.

\begin{table}[htb]
\centering
\resizebox{\columnwidth}{!}{
\begin{tabular}{c c c c}
Method & $\tau_{\rm{MgII}}$ & $\tau_{\rm{H}\beta}$ & $\tau_{\rm{H}\alpha}$ \\
\toprule
       \JAV\  $\arrvline$ & $144.7^{+24.7}_{-22.6}$ & $31.3^{+8.1}_{-4.1}$ & $27.7^{+5.3}_{-4.7}$ \\
       \PyROA\  $\arrvline$ & $157.0^{+5.4}_{-6.1}$ & $27.4^{+3.1}_{-2.7}$ & $24.6^{+2.2}_{-2.1}$ \\
\bottomrule
\end{tabular}
}
\caption{
Table comparing the 2014 Balmer and 2014-2017 \MgII\  integrated lag measurements between \JAV\ and \PyROA. We find that both lag measurement methods recover the same results within the errors.}
\label{Tab:JAV_Comparison}
\end{table}

\subsection{Measuring the Integrated Lags}
\label{Sec:IntegratedLags}
With the aim of understanding the structure of the BLR, we measure the time delays for each emission line using \PyROA. In Figure~\ref{Fig:photometric_light_curve}, we see that the there is a long period of relatively low flux (i.e., `low state') in the continuum from 2013-2019 and a rapid increase by a factor of >3 from 2019 to 2021. In our analysis, we seek to minimize the effects of non-reverberation variability of the BLR from ``line breathing'', which is characterized by the correlation between flux and BLR emission radius resulting from the dramatic increases/decreases in the continuum flux. We avoid the contribution of non-reverberation variability to the measured lags by splitting the light curves into two regions: the `low state' from 2013-2019 and the `high state' from 2022-2023.

Since \PyROA\ allows multiple responding light curves to be fit at once, we first run \PyROA\ on all the emission lines (\emissAll) for the `low state'. This allows the best-fit driving light curve model to be constrained by all of the emission lines and is presumably the best way to constrain the driving light curve if the emission lines all respond in the same qualitative way to the continuum. However, we find that \PyROA\ cannot compute good time delays while simultaneously fitting the Balmer line and \MgII\ light curves of \target. We find that the best-fit \PyROA\ driving light curve results in a poor fit to the \MgII\ light curve and even the Balmer lines have less reliable time delays due being constrained by the qualitatively different \MgII\ light curve (see Section~\ref{Sec:AnomalyMgII} for more details). We then run \PyROA\ on the Balmer lines (both \Ha\ and \Hb\ are fit simultaneously) and then \MgII\ separately. 

\begin{figure*}[t]%[ht!]
\epsscale{1.1}
\plotone{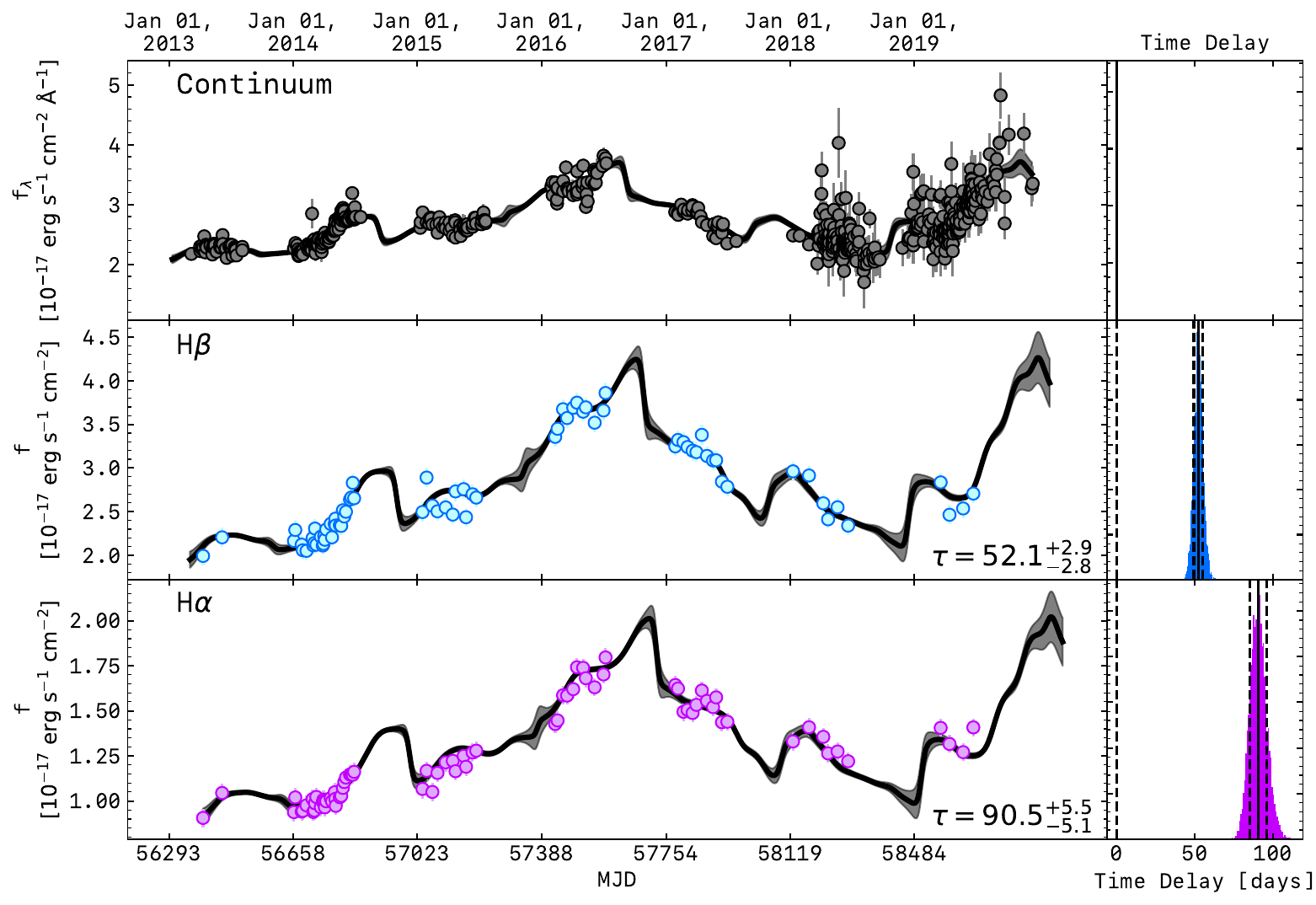}
\figcaption{`Low state' (2013-2019) Balmer light curves with best-fit \PyROA\ model overlaid as a solid black line with the corresponding error envelope. The top, middle and bottom panels show the continuum, \Hb, and \Ha\ light curves respectively. The best-fit time delay along with its corresponding errors are on the bottom right of the \Hb\ and \Ha\ light curve panels. The posterior distribution for the time delay is on the right panel of each responding light curve where the median value and 68$^{\rm th}$ percent errors are the solid and dashed lines, respectively. We also show, on the right hand panels, a vertical dashed line at the zero time delay for reference. We find an \Hb\ time delay of $52.1^{+2.9}_{-2.8}$~days and an \Ha\ time delay of $90.5^{+5.5}_{-5.1}$~days, which are longer than the time delays found in 2014 (Figure~\ref{Fig:PyROA_2014_Balmer}). This is likely due to the fact that the `low state' includes the moderate increase in flux from 2014-2017.
\label{Fig:low_state_balmer}}
\end{figure*}

\begin{figure*}[t]%[ht!]
\epsscale{1.1}
\plotone{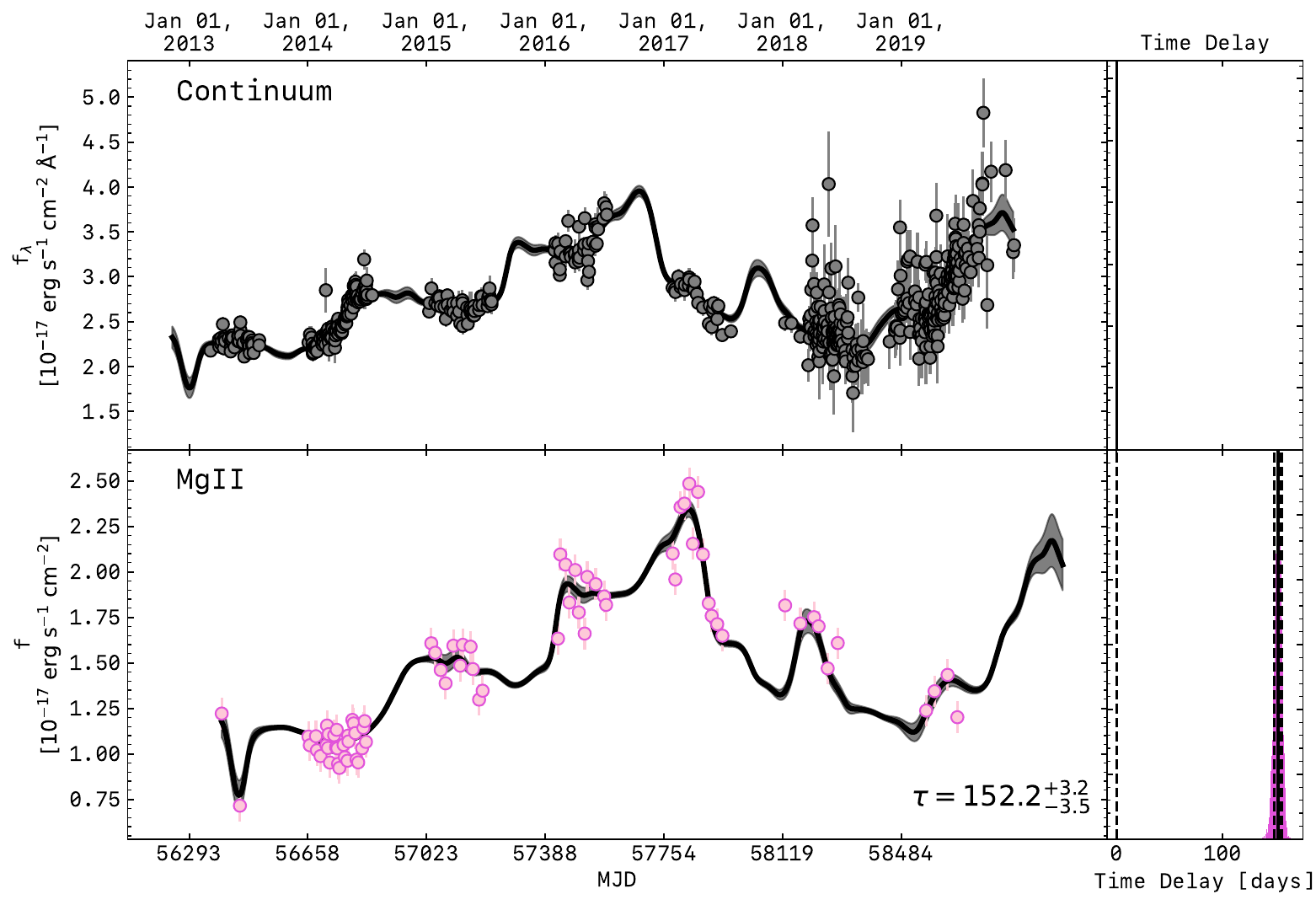}
\figcaption{`Low state' (2013-2019) \MgII\ light curve with best-fit \PyROA\ model overlaid as a solid black line with the corresponding error envelope. The top panel shows the continuum, while the bottom panel shows the \MgII\ light curve. The best-fit time delay along with its corresponding errors are on the bottom right of the \MgII\ light curve panel. The posterior distribution for the time delay is on the right panel of each responding light curve where the median value and 68$^{\rm th}$ percent errors are the solid and dashed lines, respectively. We also show, on the right hand panels, a vertical dashed line at the zero time delay for reference. We find an \MgII\ time delay of $152.2^{+3.2}_{-3.5}$~days, which is consistent with the time delay computed during the 2014-2017 time period (Figure~\ref{Fig:PyROA_2014_2017_MgII}). This is likely due to the fact that both time periods include the same range of continuum and line flux. 
\label{Fig:low_state_mgii}}
\end{figure*}

Our \PyROA\ best-fit model light curves for the `low state' are shown for the Balmer lines in Figure~\ref{Fig:low_state_balmer} and for \MgII\ in Figure~\ref{Fig:low_state_mgii}. For the Balmer lines, we find a different mean time delay for both \Ha\ and \Hb\ compared to the 2014 analysis. We find that the \Hb\ mean time delay increases by a factor of $\sim$2$\times$ and the \Ha\ mean time delay increases by a factor of $\sim$2.5$\times$. This increase is likely caused by the factor of $\sim$1.8$\times$ increase in flux from 2014 to 2016. For \MgII, we find a similar mean time delay comparing the `low state' analysis to the 2014-2017 analysis. This is due to the 2014-2017 analysis capturing the modest increase in flux, so the similar mean time delay is expected. 

\begin{figure*}[t]%[ht!]
\epsscale{1.1}
\plotone{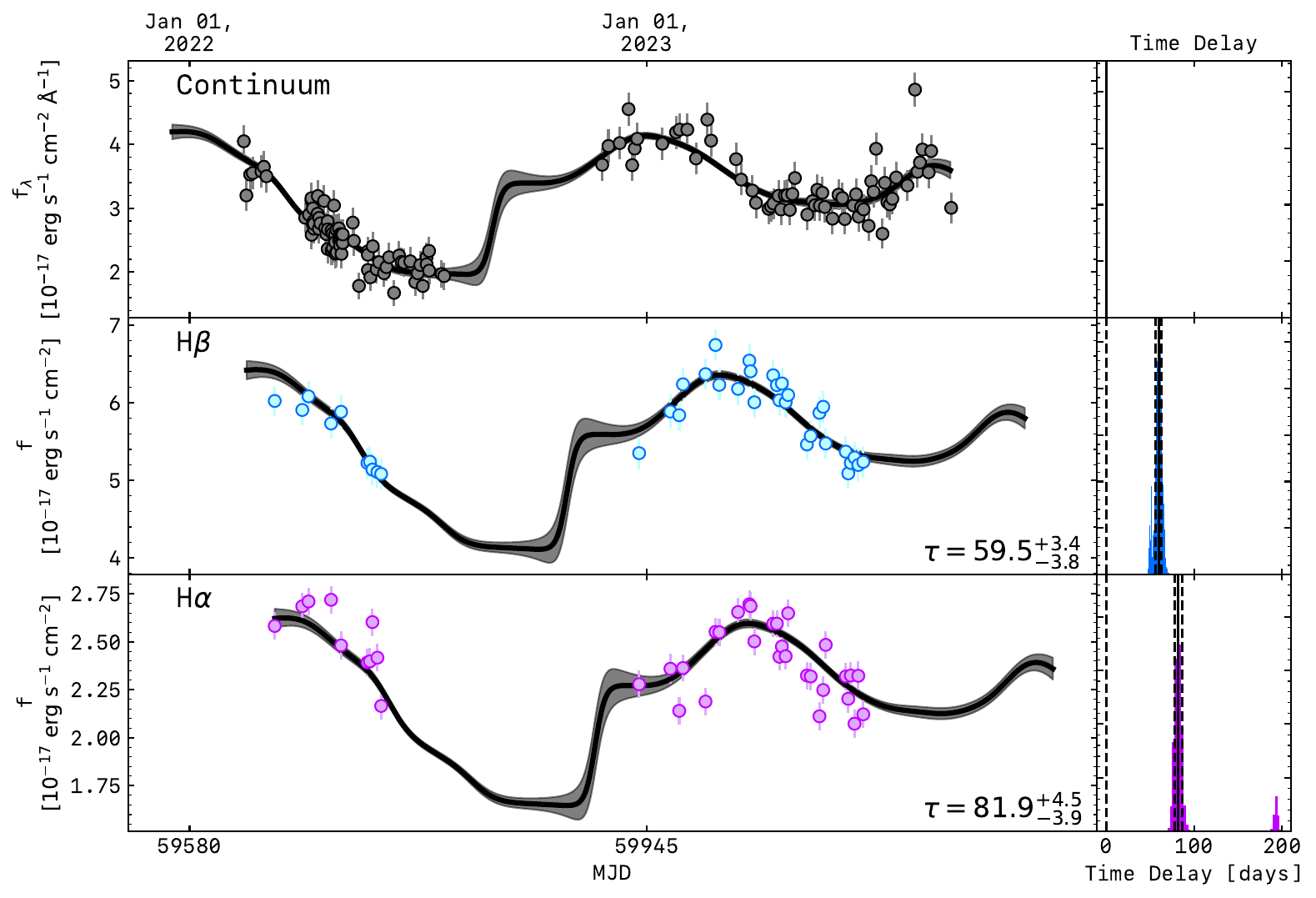}
\figcaption{`High state' (2022-2023) Balmer light curves with best-fit \PyROA\ model overlaid as a solid black line with the corresponding error envelope. The top, middle and bottom panels show the continuum, \Hb, and \Ha\ light curves respectively. The best-fit time delay along with its corresponding errors are on the bottom right of the \Hb\ and \Ha\ light curve panels. The posterior distribution for the time delay is on the right panel of each responding light curve where the median value and 68$^{\rm th}$ percent errors are the solid and dashed lines, respectively. We also show, on the right hand panels, a vertical dashed line at the zero time delay for reference. We find an \Hb\ time delay of $59.5^{+3.4}_{-3.8}$~days and an \Ha\ time delay of $81.9^{+4.5}_{-3.9}$~days. The \Hb\ time delay in the `high state' is longer than the \Hb\ time delay in the `low state'. This is consistent with the general expectation of a photoionized BLR. In other words an increase in flux will push the `optimal emission radius' (BLR radius) further out. However, we observe the opposite in the case of \Ha, where the \Ha\ `low state' has a \textit{higher} time delay than the `high state'.
\label{Fig:high_state_balmer}}
\end{figure*}

Our \PyROA\ best-fit model light curves for the `high state' are shown for the Balmer lines in Figure~\ref{Fig:high_state_balmer}. For the Balmer lines, we find a different mean time delay for both \Hb\ and \Ha\ from the 2014 analysis in \cite{Grier2017b} and a slight difference in our `low state' analysis. This is to be expected from ``line breathing'' where the increase in continuum luminosity will yield a longer mean time delay. We observe a $\sim$1.14$\times$ increase in the \Hb\ mean time delay from `low state' to `high state' resulting in a 1.61$\sigma$ difference using the average of the asymmetric error bars. While we observe a $\sim$0.09$\times$ decrease in the \Ha\ mean time delay from `low state' to `high state', resulting in a 1.27$\sigma$ difference using the average of the asymmetric error bars. This decrease in \Ha\ mean time delay from a period of low flux to high flux is unexpected from photoionization physics since a photoionization-bounded BLR should show longer lags during periods of high flux \citep{Barth2015, Wang2020}. In contrast to the Balmer lines, we find that the \MgII\ light curve is poorly constrained in the `high state' owing to a qualitatively different response to the continuum.

\subsection{The Anomalous Behavior of MgII}
\label{Sec:AnomalyMgII}
Qualitatively, the \MgII\ line appears to reverberate in a different way than the Balmer lines after the increase in received continuum luminosity in 2021. Since we are attempting to simultaneously fit all three emission-line light curves from the same driving light curve, the Balmer lines have considerably worse fits due to being constrained by the qualitatively different \MgII\ light curve. We find that the \MgII\ response is similar to the Balmer lines in the `low state', but not similar in the `high state'. We also see this in the kinematics of the lines \citep{Fries2023}.

The qualitative differences between the \MgII\ light curve and the Balmer lines could be due to the fact that the spectrophotometric reliability in the bluest wavelengths for SDSS-RM/BHM-RM is poor (see Fig. 20 in \citealt{Shen2015}). However, the \MgII\ response being similar in the `low state', but dissimilar in the `high state' suggests this is not an issue with the blue side of the spectrograph. The differences also could be due to the fact that \MgII\ is close to the dust-sublimation radius and the delay-distribution is being truncated due to dust sublimation in the outer BLR \citep{Baskin2018, Wang2020}. Another reason could be due to the fact that \MgII\ includes significant contribution from collisional excitation in addition to photoionization, which dominates for the Balmer lines \citep{Guo2020}. This added contribution of collisional excitation could lead to a BLR geometry with a broader radial extent and a more complicated response than simple line breathing. Due to these added complications, we omit the \MgII\ line for the rest of our analysis.

\begin{figure*}[t]%[ht!]
\epsscale{1.1}
\plotone{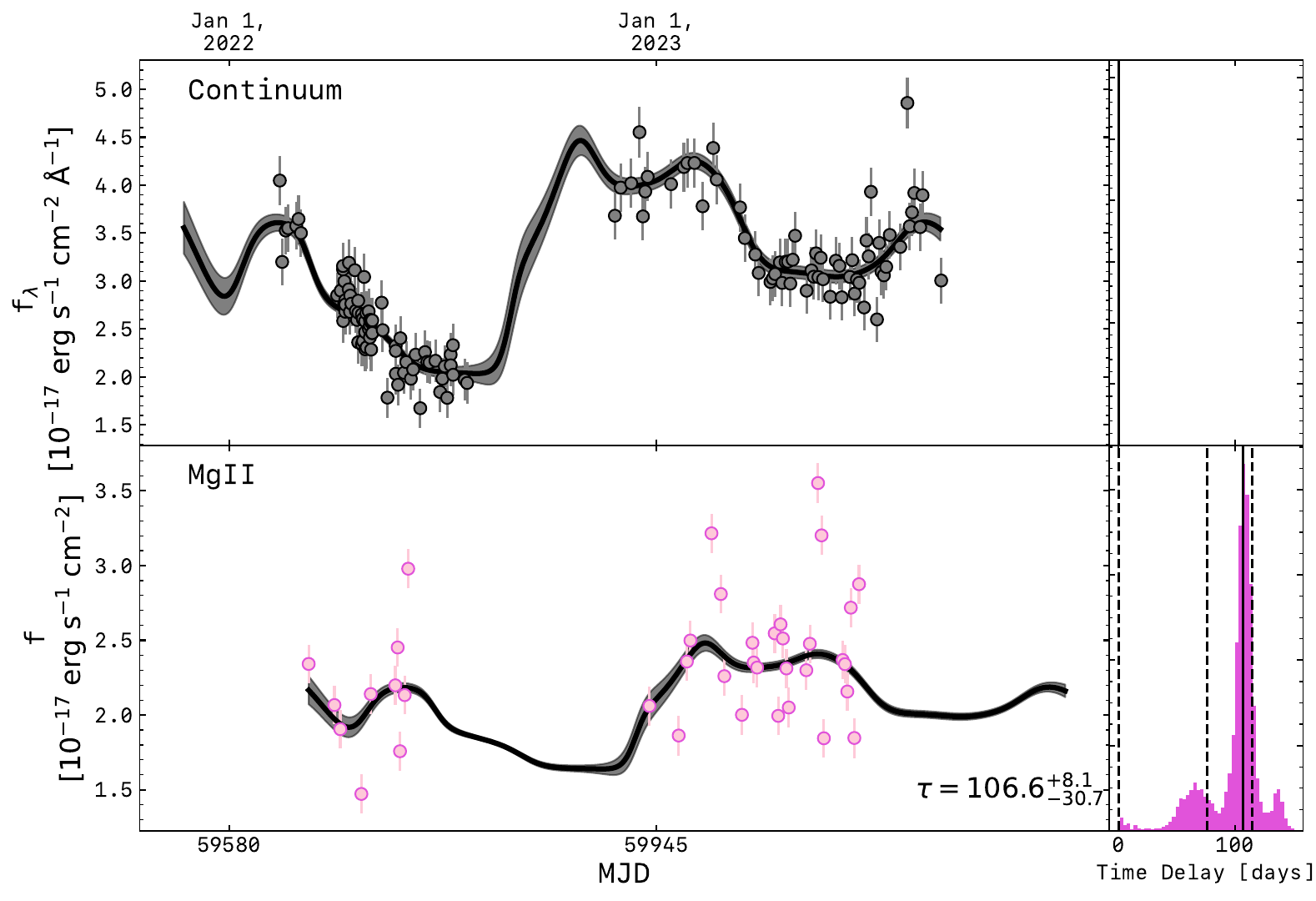}
\figcaption{`High state' (2022-2023) \MgII\ light curve with best-fit \PyROA\ model overlaid as a solid black line with the corresponding error envelope. The top and bottom panels show the continuum, \MgII\ light curves respectively. The best-fit time delay along with its corresponding errors are on the bottom right of the \MgII\ light curve panel. The posterior distribution for the time delay is on the right panel of the responding light curve where the median value and 68$^{\rm th}$ percent errors are the solid and dashed lines, respectively. We also show, on the right hand panels, a vertical dashed line at the zero time delay for reference. We find that the \MgII\ high state is not well constrained by the \PyROA\ model.
\label{Fig:high_state_mgii}}
\end{figure*}

\section{Resolving the Kinematics of the BLR}
\label{Sec4}

\subsection{Velocity-Resolved Lags}
\label{Sec:VelocityResolvedLags}
In order to probe the kinematics of the BLR, we measure the velocity-resolved response using \PyROA\ for individual velocity segments of each emission line. Following \cite{Bentz2009, Denney2009a, Grier2013, U2022}, we use two schemes to divide the emission line into 8 individual velocity segments in order to determine any potential effects of binning on the measured lags. The two schemes are as follows: (i) velocity segments of equal RMS flux and (ii) velocity segments of equal RMS velocity. Table~\ref{Tab:EqualRMSFluxBins} and Table~\ref{Tab:EqualRMSVelBins} show the breakdown for the 8 partitions for equal RMS flux and equal RMS velocity, respectively. Throughout this work, we use the equal RMS flux bins for all of our interpretations, although we confirmed that the equal RMS velocity bins exhibit the same general trends for each of the velocity-resolved lag profiles (VRLPs).

We determine the integrated flux in each partition using the method described in detail in \cite{Fries2023}. Briefly, we use a non-parametric method (See Section 3.4 of \citealt{Fries2023} for more detail) to determine the integrated broad-line flux from the continuum-subtracted and narrow-line subtracted spectra. We do this for the partitions for each emission line shown in Tables~\ref{Tab:EqualRMSFluxBins} and \ref{Tab:EqualRMSVelBins} for all epochs yielding 8 different light curves with 153 data points for each emission line for each choice of partitioning. 

For our velocity-resolved analysis, we use the same \PyROA\ priors and methodology as we did for our integrated analysis in Section~\ref{Sec:LagMeasuringMethod}. However, we simultaneously fit the velocity-resolved light curves of each emission line along with the integrated light curve resulting in 9 responding light curves in the fit. We found that fitting the velocity-resolved light curves separately and simultaneously gave the same results, and chose to fit all of them simultaneously to save computation time. Therefore, due to the velocity-resolved light curves being fit simultaneously with the integrated light curve, the integrated lags reported in this section will differ from those in Section~\ref{Sec:IntegratedLags}.

\begin{figure*}[t]%[ht!]
\epsscale{1.1}
\plotone{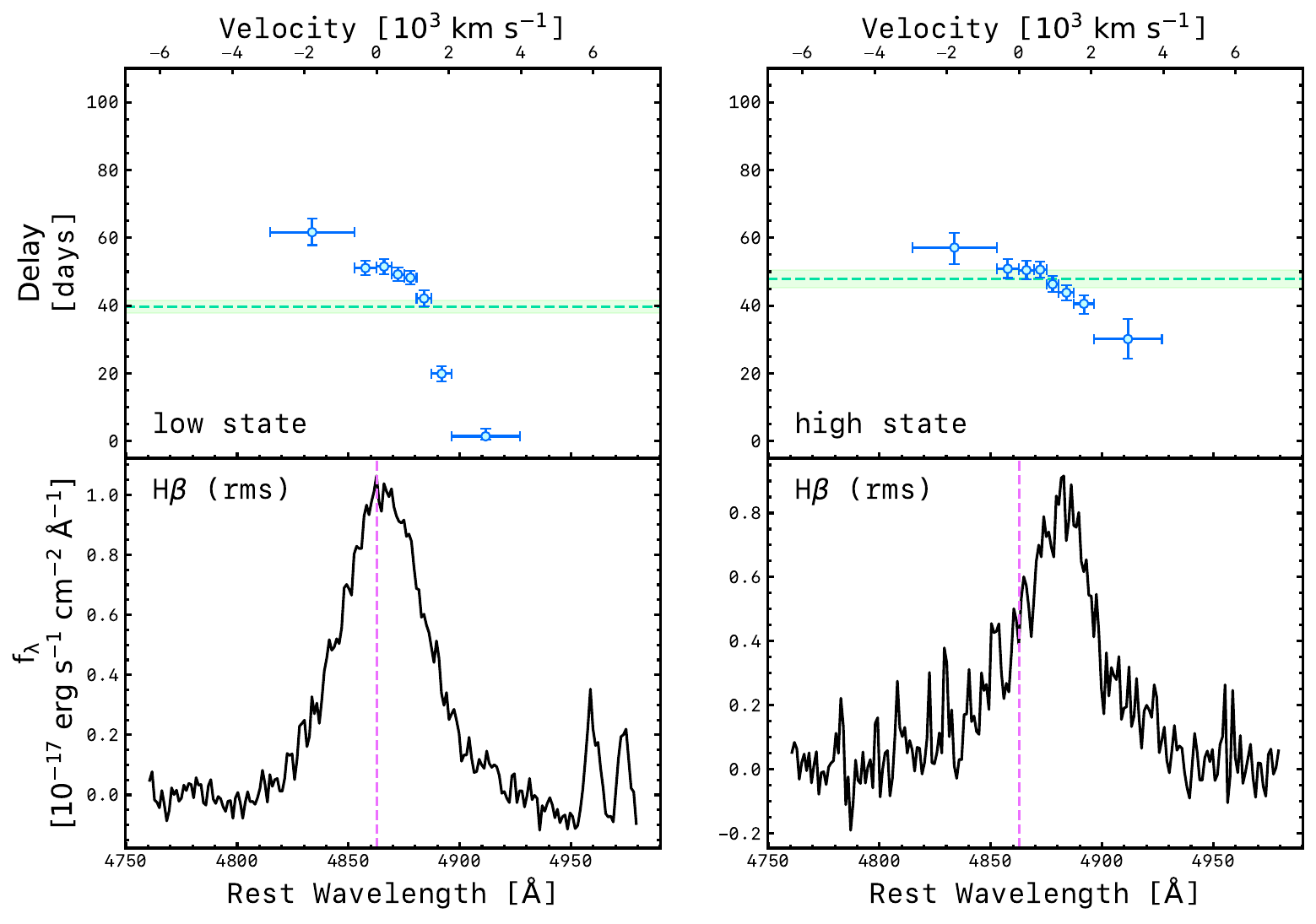}
\figcaption{VRLP for \Hb. The left side presents the lags measured from the `low state' (2013-2019) while the right side presents the lags measured from the `high state' (2022-2023). The top panels are the VRLP where the dashed green line represents the integrated time delay and the shaded green region represents $\pm$1$\sigma$ errors in the integrated time delay. For each point, the x-axis error bars represent the velocity bin as denoted in Table~\ref{Tab:EqualRMSFluxBins} and the y-axis error bars represent the error in the time delay computed from \PyROA. The bottom panels are the RMS spectrum for each state. The vertical dashed magenta line represents the systemic redshift for \Hb. The velocity-resolved lags in both the low-state and the high-state are consistent with an inflowing BLR. 
\label{Fig:Hb_evolution}}
\end{figure*}

\begin{figure*}[t]%[ht!]
\epsscale{1.1}
\plotone{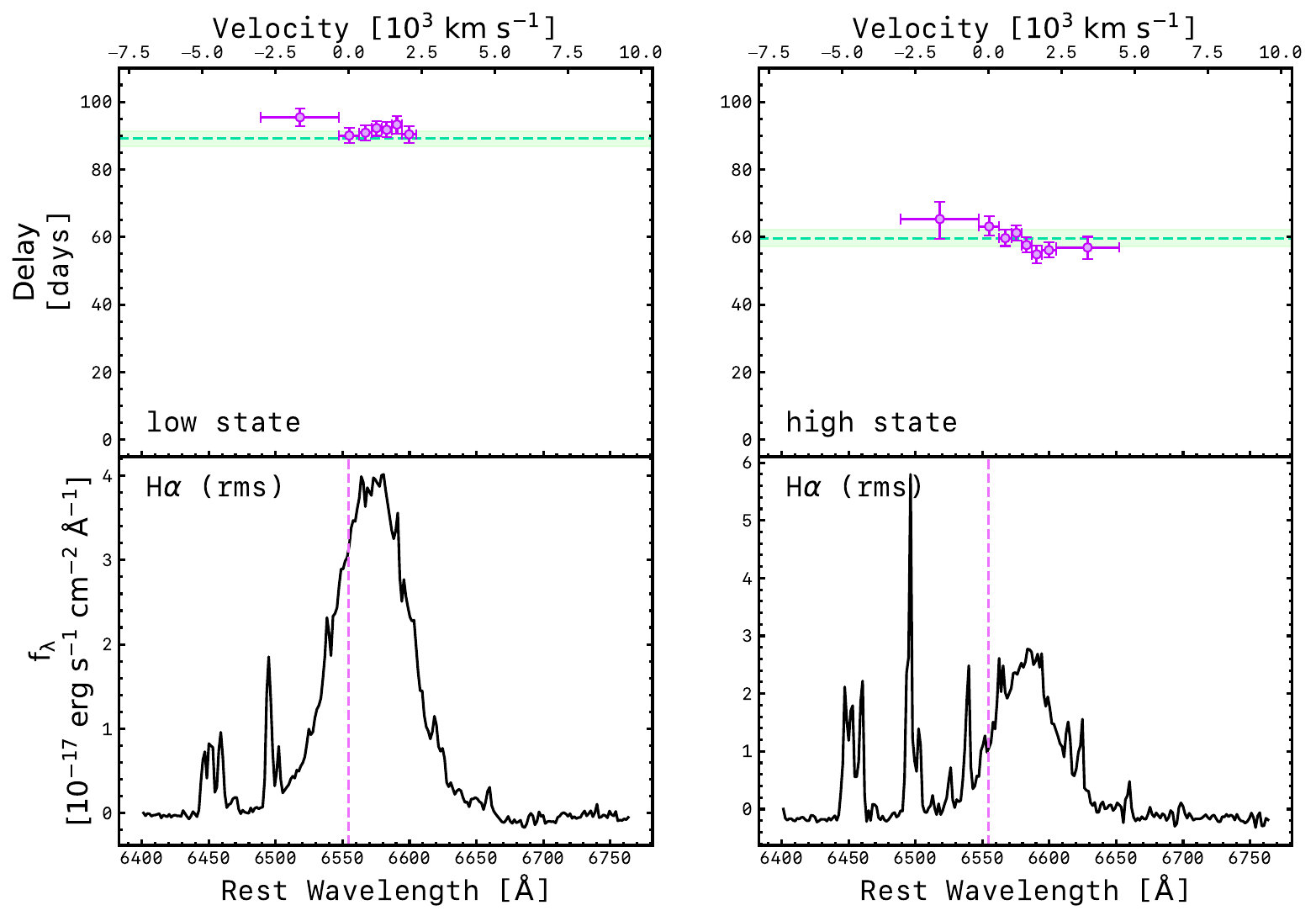}
\figcaption{VRLP for \Ha. The left side represents the `low state' (2013-2019) while the right side represents the `high state' (2022-2023). The top panels are the VRLP where the dashed green line represents the integrated time delay and the shaded green region represents $\pm$1$\sigma$ errors in the integrated time delay. For each point, the x-axis error bars represent the velocity bin as denoted in Table~\ref{Tab:EqualRMSFluxBins} and the y-axis error bars represent the error in the time delay computed from \PyROA. The bottom panels are the RMS spectrum for each state. The spikes to the left of the \Ha\ profile in both the low and high state are improperly subtracted skylines. The vertical dashed magenta line represents the systemic redshift for \Ha. The velocity-resolved lags in the low-state are consistent with a virialized BLR, while in the high-state the velocity-resolved lags are consistent with an inflowing BLR. This change in velocity-resolved lags is consistent with a kinematically stratified BLR whereby the `low state' is probing more distant, virial gas, while the `high state' is probing more nearby, inflowing gas. The integrated lags from the `high state' are roughly consistent to the integrated lag of both the low and `high state' of \Hb\ where there is also inflowing kinematics.
\label{Fig:Ha_evolution}}
\end{figure*}

We show our VRLPs for \Hb\ and \Ha\ in Figures~\ref{Fig:Hb_evolution} and~\ref{Fig:Ha_evolution}, respectively (we show the best-fit \PyROA\ light curves for our velocity-resolved analysis in Appendix~\ref{VelRM_BestFits}). Each figure shows the VRLPs for both the high and low states as denoted in Section~\ref{Sec:IntegratedLags}. Both \Ha\ and \Hb\ have high SNR light curves and we expect that the measured lag uncertainties are dominated by time resolution rather than the flux uncertainties. We note that the RMS broad line profiles of the high state \Hb\ and high and low state \Ha\ are redshifted from the systemic redshift. These radial velocity shifts for the Balmer lines are suggestive of a bulk inflow of the BLR gas with a gradient of higher velocity at smaller radii, as described in detail in \cite{Fries2023}. A VRLP of the highest time delay at the zero velocity and lowest time delays at the high velocity wings is consistent with virialized motion of the BLR. In this interpretation, the closest gas (low time delays) is moving fastest as is expected from the virial theorem (v$\sim \rm{R}^{-1/2}$). A VRLP with the longest lags in the blue wing of the emission line and shortest time delays in the red wing of the emission line indicates an infalling BLR. A VRLP with the longest lags in the red wing and the shortest lags in the blue wing indicates an outflowing BLR. A simple model for understanding these non-virial kinematics is that in the inflowing scenario the far side (longest lags) of the BLR is moving toward us (blueshifted), while the near side (shortest lags) of the BLR is moving away from us (redshifted). In the outflowing scenario the far side (longest lags) is moving away from us (redshifted), while the near side (shortest lags) is moving toward us (blueshifted) \citep{Pancoast2011, Pancoast2014}.

The \Hb\ VRLP for the `low state' has higher lags at bluer wavelengths and decreases toward redder wavelengths, which is consistent with an infalling BLR. This is consistent with the suggestion of inflow from the extremely red radial-velocity shifts in the \Hb\ broad line described by \cite{Fries2023}. The `high state' also exhibits this same infalling kinematic signature. An interesting observation is that the slope of the infalling signature from the `high state' to the `low state' seems to decrease even while both appear to be consistent with an infalling signature.

For \Ha, the `low state' has a profile that is marginally consistent with virial motion whereby the highest lags are near the line center, while the lowest lags are in the high velocity wings. However, the `high state' has a different shape and is instead marginally consistent with an inflowing BLR whereby the lowest lags are on the red side and the highest lags are on the blue side.

While \Hb\ appears to have `stable' or consistent BLR kinematics from the low-state to the high-state, \Ha\ transitions from a virialized structure in the low-state to a signature of inflow in the high-state. This suggests that there are some physical processes in the slightly outer part of the BLR changing the inferred kinematics.

\begin{table*}[t!]

\centering
\begin{tabular}{cccccccccc}
% \begin{tabular}{rrrrrrrrrr}

 & \multicolumn{3}{c}{\MgII} & \multicolumn{3}{c}{\Hb} & \multicolumn{3}{c}{\Ha} \\
  \cmidrule(lr){2-4} \cmidrule(lr){5-7} \cmidrule(lr){8-10}
 Partition &  $\langle {v} \rangle_{f}$ &  $\lambda_{\rm low}$ &  $\lambda_{\rm high}$ &  $\langle {v} \rangle_{f}$ &  $\lambda_{\rm low}$ &  $\lambda_{\rm high}$ &  $\langle {v} \rangle_{f}$ &  $\lambda_{\rm low}$ &  $\lambda_{\rm high}$\\
\toprule
         1 &            -1.98 &         3754.05 &         3793.15 &          -1.49 &       6544.85 &       6596.29 &          -1.78 &       8820.63 &       8900.19 \\
         2 &            -0.46 &         3793.15 &         3802.77 &          -0.27 &       6596.29 &       6609.97 &          -0.38 &       8900.19 &       8920.71 \\
         3 &             0.12 &         3802.77 &         3808.90 &           0.22 &       6609.97 &       6619.11 &           0.12 &       8920.71 &       8933.04 \\
         4 &             0.57 &         3808.90 &         3814.16 &           0.59 &       6619.11 &       6626.73 &           0.50 &       8933.04 &       8943.33 \\
         5 &             0.97 &         3814.16 &         3819.44 &           0.93 &       6626.73 &       6634.37 &           0.84 &       8943.33 &       8953.64 \\
         6 &             1.42 &         3819.44 &         3825.60 &           1.31 &       6634.37 &       6643.54 &           1.18 &       8953.64 &       8963.95 \\
         7 &             1.96 &         3825.60 &         3833.54 &           1.77 &       6643.54 &       6655.79 &           1.59 &       8963.95 &       8978.41 \\
         8 &             2.92 &         3833.54 &         3860.11 &           2.70 &       6655.79 &       6697.30 &           2.29 &       8978.41 &       9042.47 \\
\bottomrule
\end{tabular}
\caption{The wavelength range and characteristic velocity of the 8 equal RMS flux bins used in our velocity-resolved reverberation mapping analysis. $\langle {v} \rangle_{f}$ is the flux-weighted velocity of the bin in units of $10^{-3}~\rm{km}~\rm{s}^{-1}$, $\lambda_{\rm low}$ is the lower bound observed-frame wavelength of the bin in units of \angstrom, and $\lambda_{\rm high}$ is the upper bound observed-frame wavelength of the bin in units of \angstrom.}
\label{Tab:EqualRMSFluxBins}
\end{table*}

\begin{table*}[t!]

\centering
\begin{tabular}{cccccccccc}
% \begin{tabular}{rrrrrrrrrr}

 & \multicolumn{3}{c}{\MgII} & \multicolumn{3}{c}{\Hb} & \multicolumn{3}{c}{\Ha} \\
  \cmidrule(lr){2-4} \cmidrule(lr){5-7} \cmidrule(lr){8-10}
 Partition &  $\langle {v} \rangle_{f}$ &  $\lambda_{\rm low}$ &  $\lambda_{\rm high}$ &  $\langle {v} \rangle_{f}$ &  $\lambda_{\rm low}$ &  $\lambda_{\rm high}$ &  $\langle {v} \rangle_{f}$ &  $\lambda_{\rm low}$ &  $\lambda_{\rm high}$\\
\toprule
         1 &                -2.95 &             3760.05 &             3773.52 &              -2.48 &           6543.90 &           6563.18 &              -3.83 &           8789.64 &           8823.10 \\
         2 &                -1.86 &             3773.52 &             3787.00 &              -1.67 &           6563.18 &           6582.46 &              -2.85 &           8823.10 &           8856.56 \\
         3 &                -0.76 &             3787.00 &             3800.47 &              -0.76 &           6582.46 &           6601.74 &              -1.59 &           8856.56 &           8890.03 \\
         4 &                 0.25 &             3800.47 &             3813.94 &               0.12 &           6601.74 &           6621.02 &              -0.41 &           8890.03 &           8923.49 \\
         5 &                 1.26 &             3813.94 &             3827.42 &               0.93 &           6621.02 &           6640.30 &               0.59 &           8923.49 &           8956.95 \\
         6 &                 2.28 &             3827.42 &             3840.89 &               1.74 &           6640.30 &           6659.57 &               1.59 &           8956.95 &           8990.41 \\
         7 &                 3.29 &             3840.89 &             3854.37 &               2.63 &           6659.57 &           6678.85 &               2.63 &           8990.41 &           9023.88 \\
         8 &               4.41 &             3854.37 &   3867.84 &               3.54 &           6678.85 &  6698.13 &               3.82 &           9023.88 &           9057.34 \\
\bottomrule
\end{tabular}
\caption{The wavelength range and characteristic velocity of the 8 equal RMS velocity bins used in our velocity-resolved reverberation mapping analysis. $\langle {v} \rangle_{f}$ is the flux-weighted velocity of the bin in units of $10^{-3}~\rm{km}~\rm{s}^{-1}$, $\lambda_{\rm low}$ is the lower bound observed-frame wavelength of the bin in units of \angstrom, and $\lambda_{\rm high}$ is the upper bound observed-frame wavelength of the bin in units of \angstrom.}
\label{Tab:EqualRMSVelBins}
\end{table*}

\section{Discussion}
\label{Sec5}

\subsection{Differences in the Balmer Lines}
\label{Sec:BalmerDiff}
Figures~\ref{Fig:Hb_evolution} and~\ref{Fig:Ha_evolution} demonstrate that the inferred kinematics from our velocity-resolved reverberation mapping analysis yield different results between the Balmer lines in the `low state' despite being produced by the same emission mechanism. \Hb\ shows a signature of an infalling BLR, while \Ha\ shows the virialized signature. However, the profiles are more similar in the `high state', but there is still anomalous behavior in that the \Ha\ lag becomes shorter despite the higher continuum luminosity, while the \Hb\ lag becomes longer. Higher orders of Balmer lines (such as \Hg) for \target\ have low SNR and are unable to be used to measure velocity-resolved, as well as integrated, lags.

Both photoionization modeling and observations have shown that \Ha\ is observed from slightly further out in the BLR due to radial stratification and optical depth effects \citep{Netzer1975, Rees1989, Korista2004, Bentz2010a}. The BLR has different kinematics at different radii. The \Ha\ reverberation profile generally probes more distant gas (with potentially different kinematics) compared to the smaller-radius gas probed by \Hb. Due to this, we postulate that the different kinematic signatures at different radii in the `low state' indicates a strong kinematic gradient in the BLR gas.

\subsection{Kinematic Evolution}
\label{Sec:KinematicEvo}
Figure~\ref{Fig:Ha_evolution}, demonstrates that there is a marginal evolution of kinematics for the \Ha\ line throughout the monitoring period. In the `low state', we find a velocity-resolved response consistent with virialized motion of the BLR gas, while in the `high state' we find a velocity-resolved response indicating an infalling motion of the BLR gas. The \Hb\ response does not exhibit the same dramatic evolution, with only modest differences in the velocity-resolved lags that are consistent with inflow in both the low and high states. This modest kinematic evolution of \Ha\ could plausibly be due to some physical process within the outer BLR since \Ha\ has been shown to be slightly further out.

This evolution of the inferred \Ha\ kinematics from virial to inflowing is plausibly due to a dynamically stratified BLR. In Figure~\ref{Fig:Ha_evolution}, we see that the integrated lag during the `low state' is $\sim$90 days, while the integrated lag during the `high state' is $\sim$60 days. If we compare these integrated lags for \Ha\ to that of \Hb\ in Figure~\ref{Fig:Hb_evolution}, we see that both the `low state' and the `high state' for \Hb\ have integrated lags of $\sim$40-50 days, which is comparable to the \Ha\ `high state'. This would mean that during the `low state', \Ha\ is illuminating the virialized outer part of the BLR, while then illuminating the close, infalling gas during the `high state'. The decrease in the integrated \Ha\ lag from a period of low continuum luminosity to high continuum luminosity is somewhat paradoxical in the framework of a photoionization-bounded BLR.

% adjustment response of...
\subsection{Adjustment Time of the Broad-Line Region}
\label{Sec:AdjustmentTime}
Recent velocity-resolved reverberation mapping studies have unveiled similar evolution in the BLR kinematics. \cite{Xiao2018} used archival AGN Watch\footnote{https://www.asc.ohio-state.edu/astronomy/agnwatch/} data for NGC~5548 to produce \Hb\ velocity-delay maps with \MEM\ \citep{Horne1994}. They found an evolution of kinematics with transitions between a virial BLR and an inflowing BLR, suggesting that the shrinking of the BLR might correlate to inflow dynamics. \cite{Chen2023} studied the \Hb\ velocity-resolved response from NGC~4151 for multiple seasons over $\sim$2 decades. They found that in 4 of their seasons (1996, 1998, 2018, and 2021) the \Hb\ response was virial with possibly a contribution from inflow/outflow, while in the last season (2022) they found an outflowing BLR. They examined the velocity-resolved response with respect to the photometric light curve and found that the BLR is virial and inflowing during periods of rising continuum luminosity and that the BLR is virial and outflowing in periods of falling continuum luminosity. They posited that the rise in luminosity could be the result of enhanced accretion due to the inflowing nature of the BLR, and then once that accretion surpassed the Eddington limit the radiation pressure drives winds throughout the BLR producing an outflowing kinematic signature.

In the cases of NGC~5548 \citep{Xiao2018} and NGC~4151 \citep{Chen2023}, they measured how the radius of the BLR ($R_{\rm BLR}$) changed as a function of continuum luminosity by cross-correlating the continuum light curve and the measured BLR radius as function of time. In both cases, they found that the BLR radius changed on the timescales of $\sim$years suggesting that BLR kinematics are a combined effect of radiation pressure and virialized gas that is bound to the BH.

\begin{figure*}[t]%[ht!]
\epsscale{1.1}
\plotone{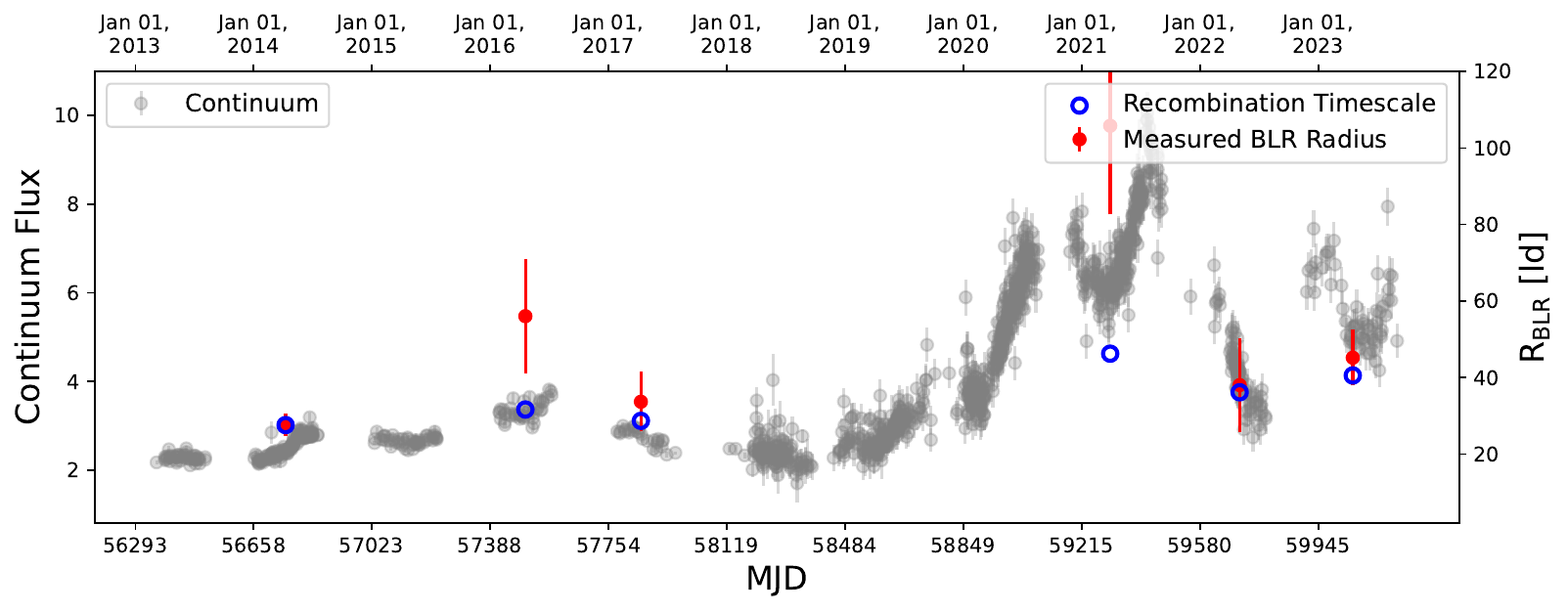}
\figcaption{Measured \Hb\ BLR radii and expected recombination timescale over plotted onto the continuum light curve. The open, blue circles show the recombination timescale expectation, the solid, red circles are the measured BLR radii, and the grey points are the continuum light curve. The points are all normalized such that they overlap in 2014. The measured BLR radii follow from our cross-correlation analysis in a given year, while the recombination timescale expectation assumes that the lag is proportional to $\sqrt{f_{\rm cont}}$. We find that the recombination timescale expectation and the measured BLR radii are the same within the uncertainties, which means the recombination timescale expectation holds for \target.
\label{Fig:RBLR_Lag}}
\end{figure*}

We test the idea of alternative mechanisms driving BLR kinematics by investigating if there is an offset between the measured BLR radii and the continuum luminosity. We have measured the BLR radius via integrated reverberation mapping in six different seasons (or years): 2014, 2016, 2017, 2021, 2022, and 2023 (the best-fit \PyROA\ light curves for these years are shown in Appendix~\ref{SeasonalLagsSec}). The BLR is assumed to respond to the central continuum on recombination timescales \citep{Krolik1991, Cackett2007}. We compare the aforementioned measured radii to the expectation of the recombination timescale in Figure~\ref{Fig:RBLR_Lag}.

If the lags and the recombination expectation line up closely, then the system is consistent with what we expect from photoionization physics. Namely, photoionization-driven optimal emitting regions of the BLR with a short recombination timescale. A scenario where the BLR is flowing in and out due to radiative winds would instead have an ``adjustment delay'', of the order of the dynamical time, between the expectation value and the corresponding BLR radius.

We find that our measured BLR do not significantly differ from a recombination timescale expectation of BLR variations and we do not see an ``adjustment time'' to the measured BLR radii, in contrast to the systems studied by \cite{Xiao2018} and \cite{Chen2023}.

\subsection{Implications for Black Hole Mass in High Redshift QSOs}
\label{Sec:BH_MassImp}
The foundation of black hole mass estimation beyond the local Universe rests on the assumption that the BLR is moving in orbits dominated by gravity (i.e., virial motion). This is true both for masses from reverberation mapping, and for the ``single-epoch'' masses based on the radius-luminosity relation \citep{Bentz2013} that is calibrated by reverberation mapping. Black hole mass estimates via reverberation mapping are found using:

\begin{equation}
    \label{RM_Mass}
    M = f \frac{v^{2} R_{\rm BLR}}{G}
\end{equation}
where $v$ is the width of a particular broad emission line, $R_{\rm{BLR}}$ is the ``optimal emission radius'' of that line using $R_{\rm BLR} = c \tau$, $G$ is the gravitational constant, and $f$ is a dimensionless factor that is introduced to compensate for our ignorance of the BLR kinematics, geometry, and orientation.

The VRLPs from \target\ indicate that the \Hb\ low and high states are inflowing as well as the \Ha\ `high state'. This presence of non-virial kinematics within the BLR of \target\ puts pressure on the foundation of black hole mass estimation. Moreover, even further pressure is put on the foundation of black hole mass estimation by the observation that we see the inferred \Ha\ kinematics change from one state to the next in \target.

\begin{figure*}[t]%[ht!]
\epsscale{1.1}
\plotone{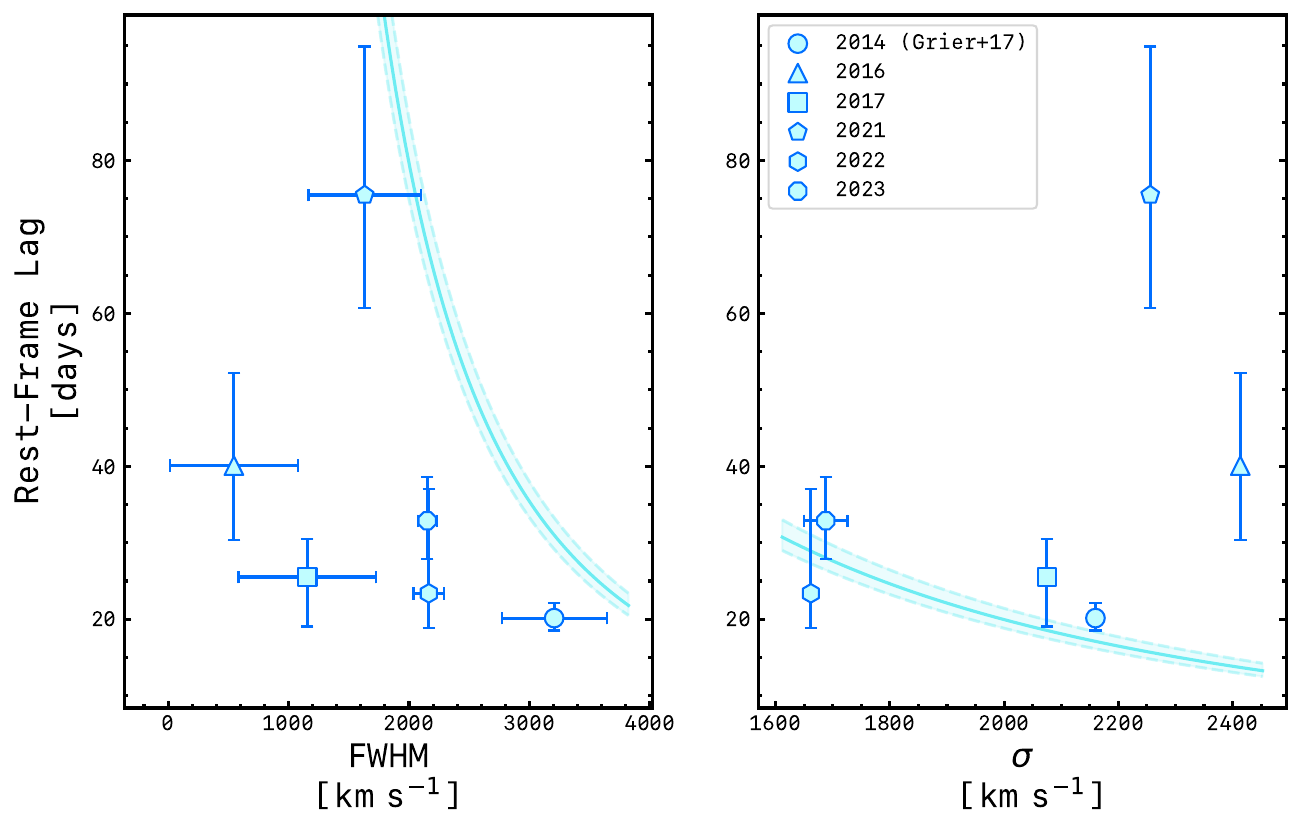}
\figcaption{Rest-frame lag vs. line width (FWHM and $\sigma$) for \Hb\ in \target. The left panel shows the rest-frame lag vs. FWHM and the right panel shows the rest-frame lag vs. $\sigma$. Different marker shapes indicate different monitoring periods as indicated by the legend. In both panels, the solid blue line indicates the line of constant \Hb-derived mass (indicating the virial product) from \cite{Grier2017b} with the shaded blue region being the errors in the \Hb\ mass. We see that the measurements of FWHM and $\sigma$ for \Hb\ do not follow the line of constant mass and therefore the virial product is not constant. This indicates that the virial product is not constant, indicating non-virial and variable BLR kinematics such that the estimated BH mass could plausibly be incorrect.
\label{Fig:HbVirialProd}}
\end{figure*}

\begin{figure*}[t]%[ht!]
\epsscale{1.1}
\plotone{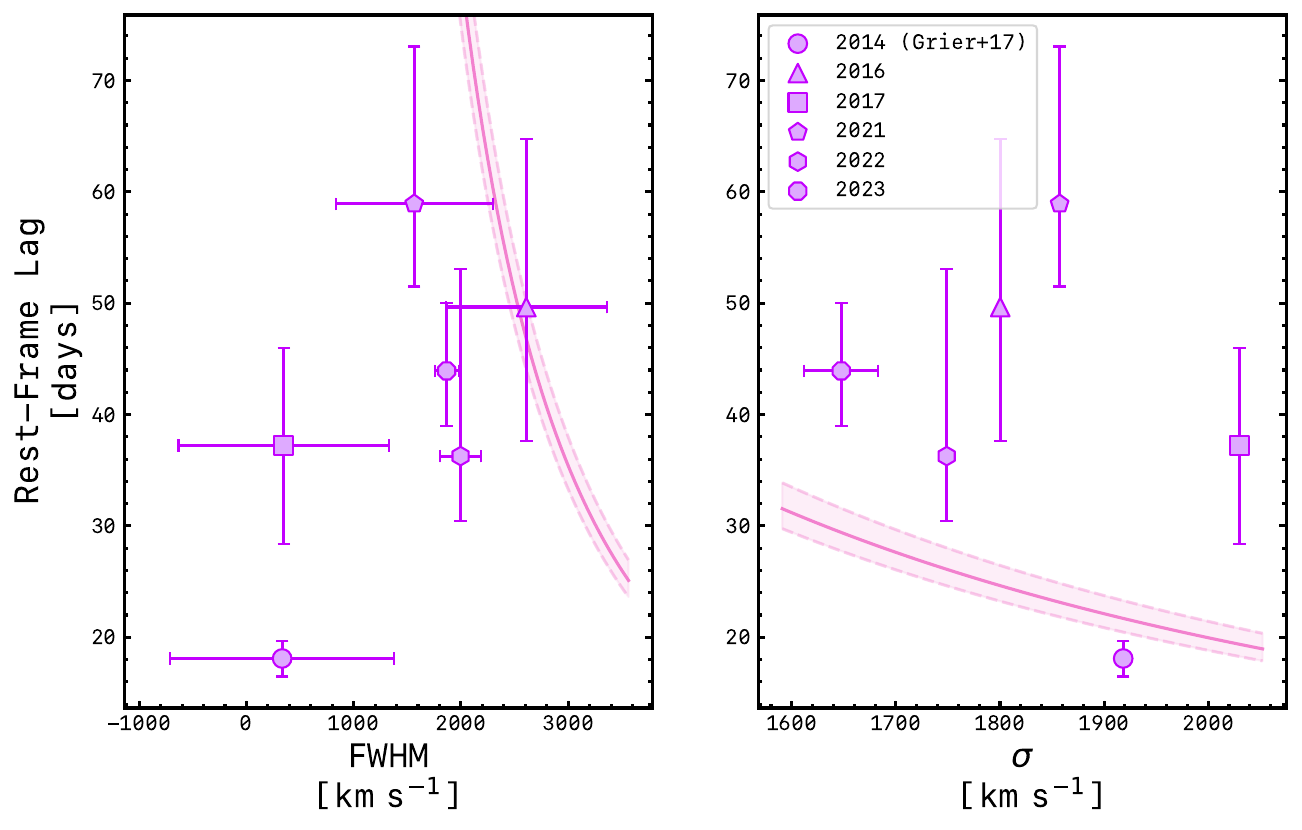}
\figcaption{Rest-frame lag vs. line width (FWHM and $\sigma$) for \Ha\ in \target. The left panel shows the rest-frame lag vs. FWHM and the right panel shows the rest-frame lag vs. $\sigma$. Different marker shapes indicate different monitoring periods as indicated by the legend. In both panels, the solid purple line indicates the line of constant \Ha-derived mass (indicating the virial product) from \cite{Grier2017b} with the shaded purple region being the errors in the \Ha\ mass. We see that the measurements of FWHM and $\sigma$ for \Ha\ do not follow the line of constant mass and therefore the virial product is not constant. This indicates that in the presence of non-virial and changing BLR kinematics, the virial product fails to be constant and the estimated BH mass could plausibly be incorrect.
\label{Fig:HaVirialProd}}
\end{figure*}

In order to understand if the changing kinematics of BLR affect the mass measurement of \target, we test the stability of the virial product. In Equation~\ref{RM_Mass}, the virial product is the $v^{2}~R_{\rm BLR}/G$ term. This term should be constant such that $v^{2}$ and $R_{\rm BLR}$ essentially compensate for each other (i.e., if $R_{\rm BLR}$ increases, $v^{2}$ decreases). In Figures~\ref{Fig:HbVirialProd} and \ref{Fig:HaVirialProd}, we plot the virial product (for both FWHM and $\sigma$ measured from the RMS spectrum) for 6 different time periods (the best-fit \PyROA\ light curves for these years are shown in Appendix~\ref{SeasonalLagsSec}). In both Figures~\ref{Fig:HbVirialProd} and \ref{Fig:HaVirialProd}, the solid line indicates the line of constant mass from \cite{Grier2017b} (i.e., points lying on this line would have a constant virial product) and the shaded region represents the errors in the mass. We see that for both \Hb\ and \Ha, both the FWHM and $\sigma$ do not follow this line of constant mass and therefore the virial product is not constant in \target. The BLR radius and velocity (from both FWHM and $\sigma$) \textit{do not obey a constant virial product}.

Figures~\ref{Fig:HbVirialProd} and \ref{Fig:HaVirialProd} demonstrate that the virial product of \target\ is not constant over the entire monitoring period of 2013-2023. The variable BLR kinematics suggest that the basic assumption behind BH mass, a virial BLR bound to the central black hole, is fundamentally flawed. As mentioned in \cite{Fries2023}, \target\ was chosen for its unusual line-profile variability, but these observations do not exist for most AGN (and despite the difficulties of SNR, cadence, and duration, there are two similar AGN reported in the literature \citep{Xiao2018, Chen2023}. It is possible that luminous quasars, probed by SDSS-RM and BHM-RM but unexplored by previous decades of single-object RM studies, are more likely to have non-virial BLR kinematics due to larger radiation pressure. More studies of luminous quasars with velocity-resolved RM are needed to understand if the basic assumptions of BH masses are valid for all AGN. 

\section{Conclusions}
\label{Sec6}
We have presented velocity-resolved reverberation mapping of the luminous quasar \target. We split the light curves into two states corresponding with the 3x increase in flux: the `low state' from 2013-2019 and the `high state' from 2022-2023. We find that the inferred kinematics from the velocity-resolved analysis are different between the Balmer lines and, more so for \Ha, they change from `low state' to `high state'. We find that changes in the BLR radii correspond to the continuum luminosity variations of the quasar, consistent with a short recombination timescale and not requiring a longer dynamical adjustment time associated with the restructuring of the BLR gas. We also find that the virial product throughout the monitoring period is non-constant, offering many questions for the enterprise of black-hole mass measurements.

We interpret the differences between the Balmer lines as a product of radial stratification in the BLR where, due to optical depth effects, the \Ha\ responding region is further out from the \Hb\ responding region. This implies that the BLR kinematics in \target\ are stratified as well such that the kinematics are not ubiquitous throughout the BLR. 

We interpret the inferred kinematic evolution of \Ha\ as kinematic stratification in the BLR, whereby the part of the BLR probed by the \Ha\ low-state is further from the black hole and virial. Conversely, in the high-state the \Ha\ part of the BLR being probed is closer in and inflowing, which agrees with the \Hb\ radii and kinematics for both states.

We examine the time delay between changes in the BLR radii and luminosity and find that the time delay is of the same magnitude as the recombination timescale unlike some recent velocity-resolved reverberation mapping studies \citep{Xiao2018, Chen2023}. This effectively tests, and verifies, the assumption of BLR variation on recombination timescales for this object.

Finally, we examine the stability of the virial product in \target. We find that the virial-product, in both FWHM and $\sigma$, for the Balmer lines is not constant (i.e., they do not follow the line of constant mass). This is plausibly due to the non-virial and changing kinematics inferred from our velocity-resolved reverberation mapping analysis. The non-virial nature of a luminous quasar like \target\ could have broad implications for BH mass estimates in high redshift quasars. If there is a significant non-virial contribution to the BLRs kinematics, then the masses will be dramatically overestimated due to the added velocity from non-virial kinematics.

Since this study only examines a single object, we do not have a good sense for how rare (or common) the inconsistent virial product phenomenon is. Most studies investigating BLR kinematics have used low redshift Seyfert 1 AGN \citep{Bentz2009, Barth2011, Du2016, U2022}, but our understanding of black hole growth over cosmic time relies heavily on our knowledge of luminous quasars. This study does represent a caution for the interpretation of black hole masses for individual luminous quasars unless the stability of the virial product can be shown through time domain spectroscopy. Unfortunately, this is challenging for high redshift quasars because of cosmological time dilation. 

Instead, we might rely on the line profile. In the case of \target, the broad emission lines are significantly redshifted compared to the systemic narrow lines and the broad emission line profiles are significantly boxier than Gaussian \citep{Fries2023}. The shape of the broad emission line profile was also identified by \cite{Villafana2022} as a possible contributor to spectroscopic mass measurement values. Considering recent JWST observations of overmassive black holes in the early Universe \citep{Harikane2023, Maiolino2023, Paccuci2023}, we suggest caution in black-hole mass estimates in high redshift quasars as non-virial kinematics could be dramatically overestimating the masses of these black holes. More studies of velocity-resolved lags in luminous QSOs are needed to determine if the variable kinematics of \target\ are common or unusual.

\section{Acknowledgements}
LBF, JRT, and MCD acknowledge support from NSF grant CAREER-1945546, and with CJG acknowledge support from NSF grant AST-2108668. JRT, CJG, and YS also acknowledge support from NSF grants AST-2009539 and AST-2009947. MK acknowledges support by DFG grant KR 3338/4-1. B.T. acknowledges support from the European Research Council (ERC) under the European Union's Horizon 2020 research and innovation program (grant agreement 950533) and from the Israel Science Foundation (grant 1849/19). X.L. acknowledges support from NSF grant AST-2206499. CR acknowledges support from the Fondecyt Iniciaci\'on grant 11190831 and ANID BASAL project FB210003. RJA was supported by FONDECYT grant number 1231718 and by the ANID BASAL project FB210003. M. L. M.-A. acknowledges financial support from Millenium Nucleus NCN$19\_058$ (TITANs). P. B. H acknowledges support from NSERC grant 2023-05068.

Funding for the Sloan Digital Sky Survey V has been provided by the Alfred P. Sloan Foundation, the Heising-Simons Foundation, the National Science Foundation, and the Participating Institutions. SDSS acknowledges support and resources from the Center for High-Performance Computing at the University of Utah. SDSS telescopes are located at Apache Point Observatory, funded by the Astrophysical Research Consortium and operated by New Mexico State University, and at Las Campanas Observatory, operated by the Carnegie Institution for Science. The SDSS web site is \url{www.sdss.org}.

SDSS is managed by the Astrophysical Research Consortium for the Participating Institutions of the SDSS Collaboration, including Caltech, The Carnegie Institution for Science, Chilean National Time Allocation Committee (CNTAC) ratified researchers, The Flatiron Institute, the Gotham Participation Group, Harvard University, Heidelberg University, The Johns Hopkins University, L'Ecole polytechnique f\'{e}d\'{e}rale de Lausanne (EPFL), Leibniz-Institut f\"{u}r Astrophysik Potsdam (AIP), Max-Planck-Institut f\"{u}r Astronomie (MPIA Heidelberg), Max-Planck-Institut f\"{u}r Extraterrestrische Physik (MPE), Nanjing University, National Astronomical Observatories of China (NAOC), New Mexico State University, The Ohio State University, Pennsylvania State University, Smithsonian Astrophysical Observatory, Space Telescope Science Institute (STScI), the Stellar Astrophysics Participation Group, Universidad Nacional Aut\'{o}noma de M\'{e}xico, University of Arizona, University of Colorado Boulder, University of Illinois at Urbana-Champaign, University of Toronto, University of Utah, University of Virginia, Yale University, and Yunnan University.

LBF is deeply indebted to Bren Backhaus and Jonathan Mercedes Feliz for their invaluable scientific and data visualization discussions.

\software{{\tt AstroPy} \citep{Astropy2013, Astropy2018},
{\tt Matplotlib} \citep{Matplotlib2007}, 
{\tt NumPy} \citep{NumPy2020}, {\tt SciPy} \citep{scipy2020}}, \PyROA\ \citep{Donnan2021}

\begin{appendix}

\section{Velocity-Resolved \PyROA\ Fits}
\label{VelRM_BestFits}
\begin{figure*}[t]%[ht!]
\epsscale{1.1}
\plotone{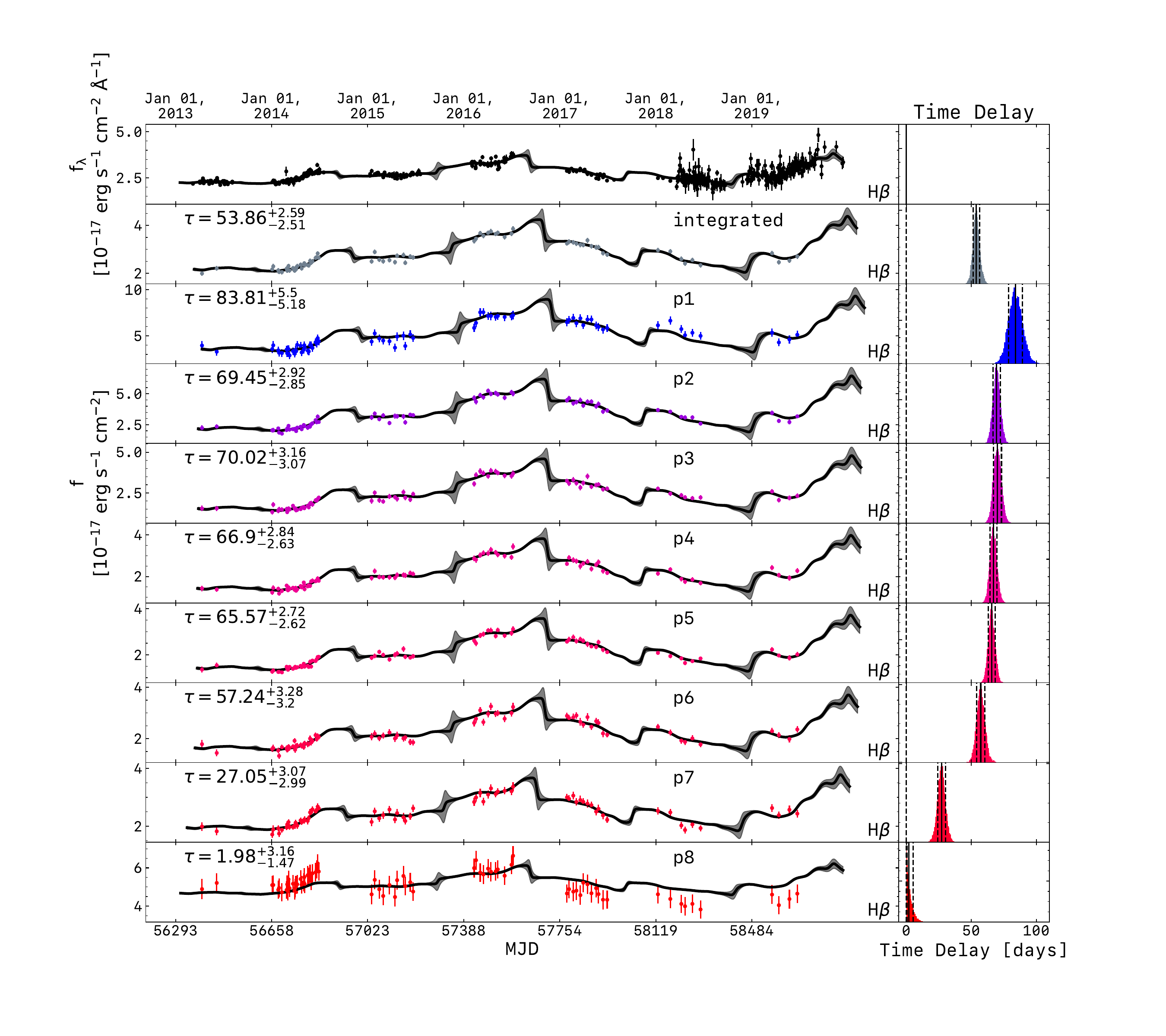}
\figcaption{`Low state' \Hb\ velocity-resolved light curves with best-fit \PyROA\ model overlaid as a solid black line with the corresponding error envelope. The top panel shows the continuum light curve, the panel below that shows the integrated light curve, and subsequent light curves are the partitioned light curves (colored from blue to red). Each partitioned light curve is labeled "p\#", where \# is the partition number outlined in Table~\ref{Tab:EqualRMSFluxBins}. The best-fit time delays along with their corresponding errors are shown in the top left of each light curve. The posterior distribution for the time delay is on the right panel of each responding light curve where the median value and 68$^{\rm th}$ percent errors are the solid and dashed lines, respectively.
\label{Fig:HbVelRM_delagged_low_state}}
\end{figure*}

\begin{figure*}[t]%[ht!]
\epsscale{1.1}
\plotone{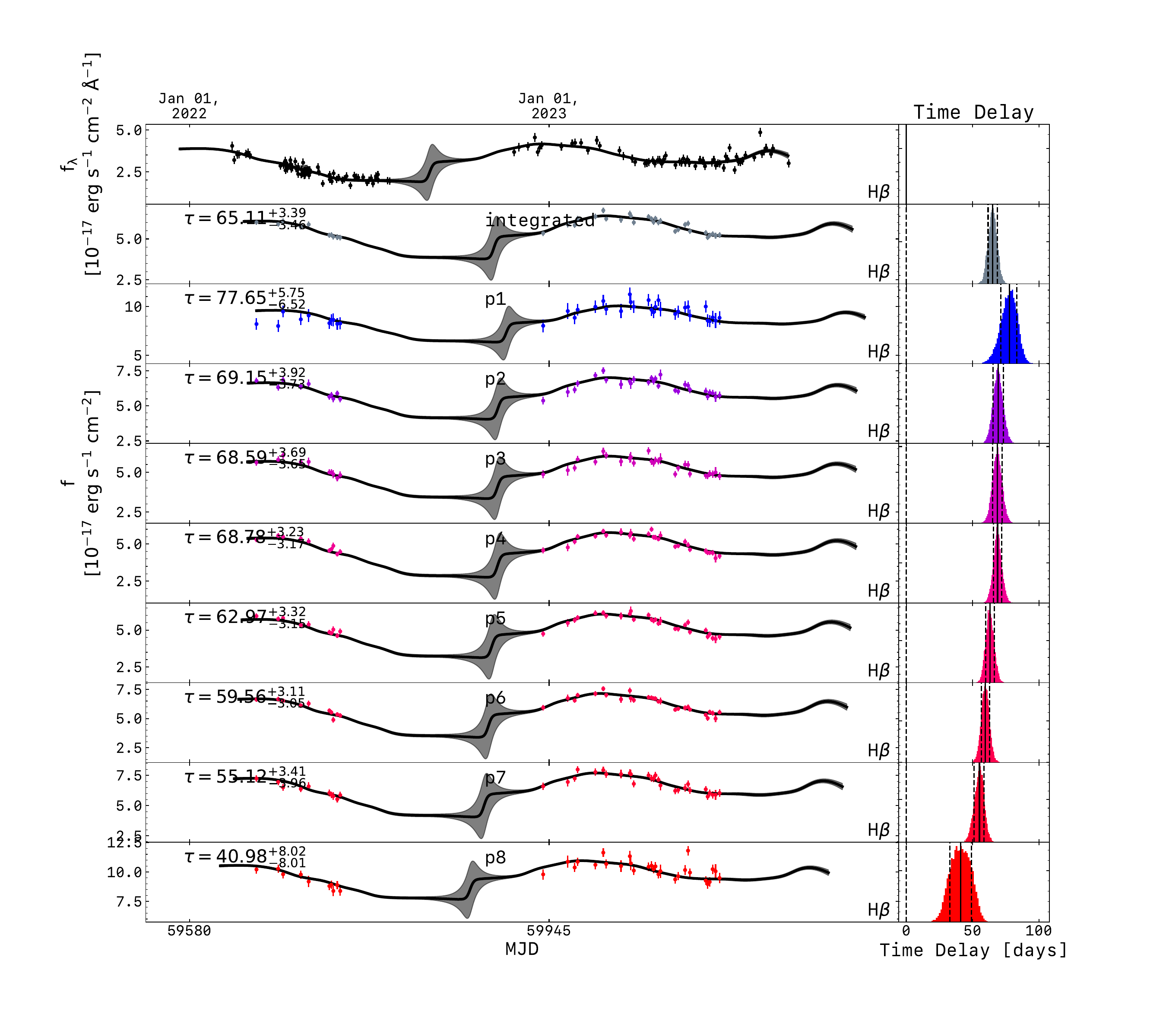}
\figcaption{`High state' \Hb\ velocity-resolved light curves with best-fit \PyROA\ model overlaid as a solid black line with the corresponding error envelope. The top panel shows the continuum light curve, the panel below that shows the integrated light curve, and subsequent light curves are the partitioned light curves (colored from blue to red). Each partitioned light curve is labeled "p\#", where \# is the partition number outlined in Table~\ref{Tab:EqualRMSFluxBins}. The best-fit time delays along with their corresponding errors are shown in the top left of each light curve. The posterior distribution for the time delay is on the right panel of each responding light curve where the median value and 68$^{\rm th}$ percent errors are the solid and dashed lines, respectively.
\label{Fig:HbVelRM_delagged_high_state}}
\end{figure*}

\begin{figure*}[t]%[ht!]
\epsscale{1.1}
\plotone{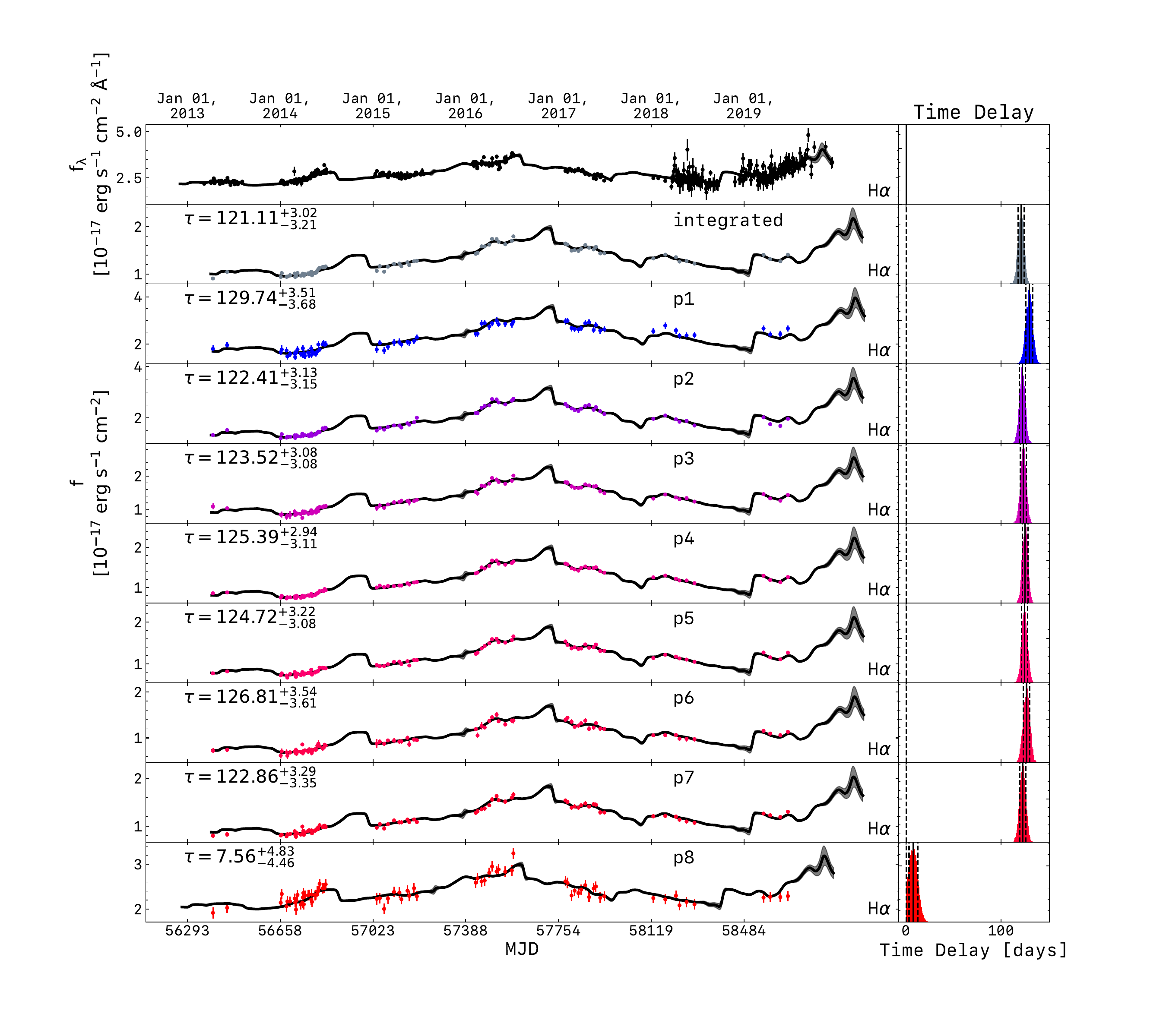}
\figcaption{Low state \Ha\ velocity-resolved light curves with best-fit \PyROA\ model overlaid as a solid black line with the corresponding error envelope. The top panel shows the continuum light curve, the panel below that shows the integrated light curve, and subsequent light curves are the partitioned light curves (colored from blue to red). Each partitioned light curve is labeled "p\#", where \# is the partition number outlined in Table~\ref{Tab:EqualRMSFluxBins}. The best-fit time delays along with their corresponding errors are shown in the top left of each light curve. The posterior distribution for the time delay is on the right panel of each responding light curve where the median value and 68$^{\rm th}$ percent errors are the solid and dashed lines, respectively.
\label{Fig:HaVelRM_delagged_low_state}}
\end{figure*}

\begin{figure*}[t]%[ht!]
\epsscale{1.1}
\plotone{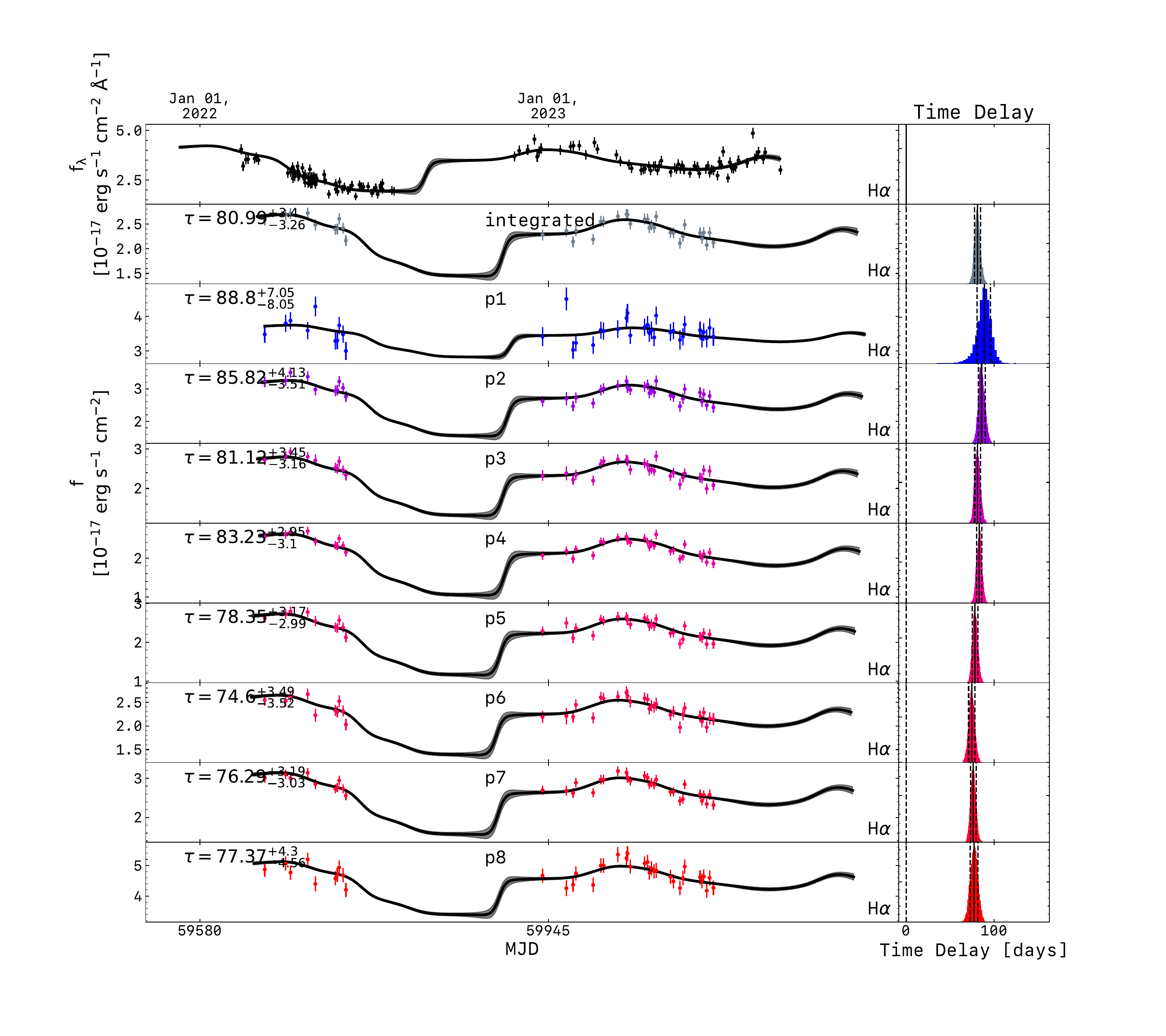}
\figcaption{`High state' \Ha\ velocity-resolved light curves with best-fit \PyROA\ model overlaid as a solid black line with the corresponding error envelope. The top panel shows the continuum light curve, the panel below that shows the integrated light curve, and subsequent light curves are the partitioned light curves (colored from blue to red). Each partitioned light curve is labeled "p\#", where \# is the partition number outlined in Table~\ref{Tab:EqualRMSFluxBins}. The best-fit time delays along with their corresponding errors are shown in the top left of each light curve. The posterior distribution for the time delay is on the right panel of each responding light curve where the median value and 68$^{\rm th}$ percent errors are the solid and dashed lines, respectively.
\label{Fig:HaVelRM_delagged_high_state}}
\end{figure*}

We present our velocity-resolved \PyROA\ best-fit models here in Figs~\ref{Fig:HbVelRM_delagged_low_state}, \ref{Fig:HbVelRM_delagged_high_state}, \ref{Fig:HaVelRM_delagged_low_state}, and \ref{Fig:HaVelRM_delagged_high_state} that are discussed in Section~\ref{Sec:VelocityResolvedLags}.

\section{Virial Product Seasonal \PyROA\ Fits}
\label{SeasonalLagsSec}

\begin{figure*}[t]%[ht!]
\epsscale{1.1}
\plotone{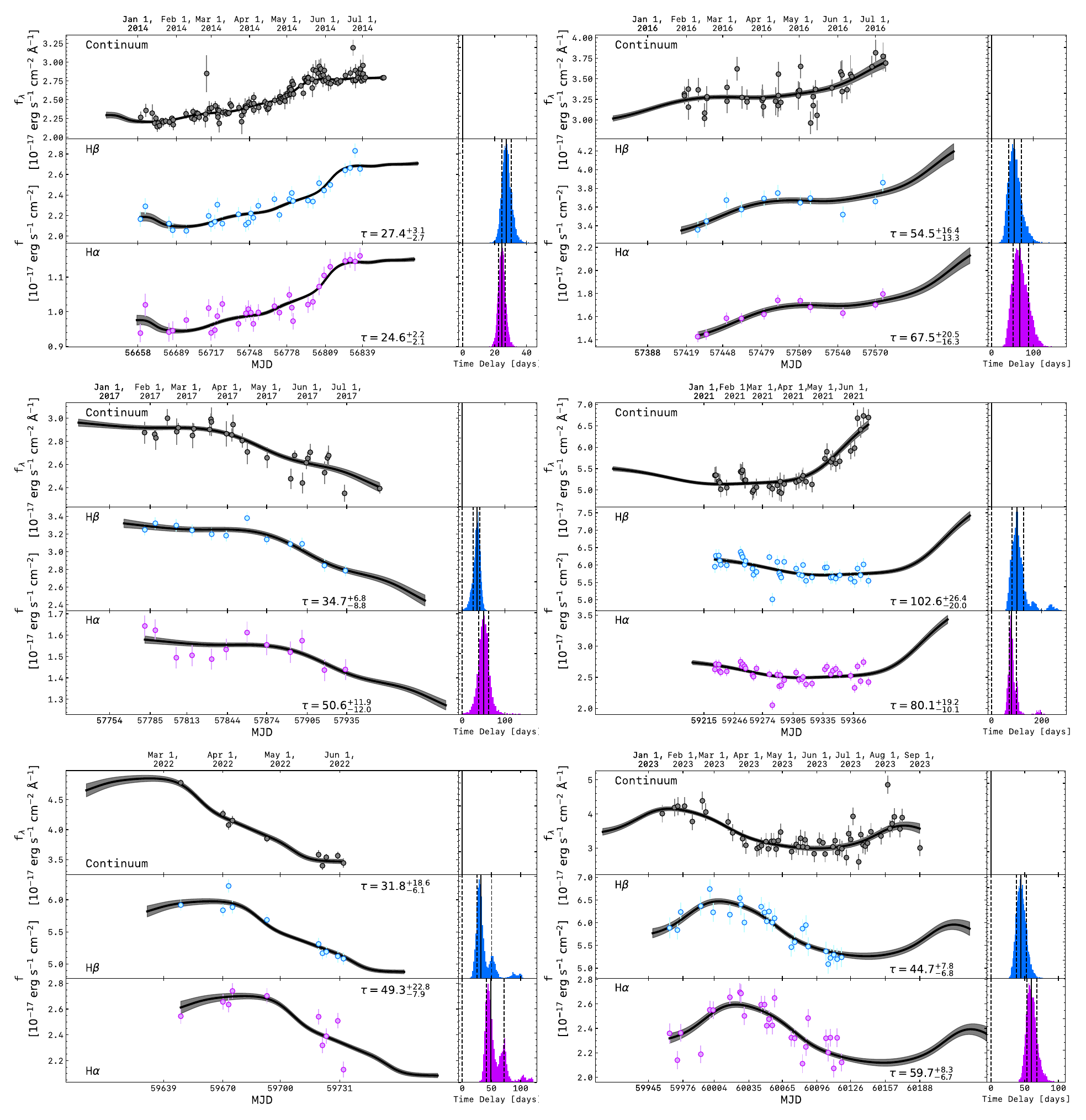}
\figcaption{Best-fit \PyROA\ light curves for the seasons shown in Figures~\ref{Fig:RBLR_Lag}, \ref{Fig:HbVirialProd}, \ref{Fig:HaVirialProd}. Years 2014 and 2016, 2017 and 2021, and 2022 and 2023 are shown from left to right in the top, middle, and bottom panels, respectively.
\label{Fig:SeasonalLags}}
\end{figure*}

We present our seasonal integrated lag measurements for the years 2014, 2016, 2017, 2021, 2022, and 2023 in Figure~\ref{Fig:SeasonalLags}. These lags are the measurements shown in Figures~\ref{Fig:RBLR_Lag}, \ref{Fig:HbVirialProd}, and \ref{Fig:HaVirialProd}.

\end{appendix}

\newpage{}

\bibliography{lib}{}

\end{document}